# Beyond costs: Mapping Norwegian youth preferences for a more inclusive energy transition

Muhammad Shahzad Javed[1,*], Karin Fossheim[2], Paola Velasco-Herrejón[1], Nikolai Elias Koop[3], Matylda N. Guzik[1], Charles Dana Samuelson[2], Beate Seibt[2], Marianne Zeyringer[1]

[1]Department of Technology Systems, University of Oslo, Norway.

[2]Department of Psychology, University of Oslo, Oslo, Norway.

[3]Department of Mathematics, University of Oslo, Oslo, Norway.

[*]Corresponding author(s). E-mail(s): m.s.javed@its.uio.no

## Abstract

Environmental movements and climate strikes have made it apparent that youth feel excluded from the ongoing energy transformation process, highlighting the crucial need for their engagement to achieve a socially accepted transition. This interdisciplinary study focuses on the Norwegian electricity system and involves conducting educational workshops with high school students aged 15 to 16 to ascertain their perspectives and socio-techno-economic preferences towards a net-zero energy system. The workshops were structured into three segments, starting with the dissemination of common knowledge about energy and climate, followed by segments of interactive activities designed to explore and develop a shared understanding of various aspects of energy transition. Three rounds of questionnaires, administered at distinct time intervals, assessed changes in students' attitudes and socio-techno-economic preferences. Our findings show that 33% of pupils favored exclusively offshore wind as a main energy source, while 35% opted to combine it with solar energy, indicating that over 68% of participants viewed offshore wind as a favorable solution. Although 32% supported some form of land-based wind turbines, there was strong disagreement about having wind parks in agricultural, forested, and residential areas. Preferences also exhibited considerable regional variation; solar installations were favored in southern and southeastern Norway, while wind farms were suggested for central and northern regions. Pupils emphasized energy independence, showed reluctance towards demand response (i.e., adjusting energy use), prioritized reducing carbon emissions and preserving biodiversity over minimizing electricity costs. Despite cost-minimization being at the core of most existing and broadly utilized energy system models, youth deemed it the least important factor, highlighting a disconnect between traditional modeling priorities and their perspectives. Thus, integrating the perspectives of young people opens space for energy modelers to explore uncertainties related to technological adoption by generating socially relevant scenarios.

## 1. Introduction

Energy system models are widely used to advise policymakers and planners on the best strategies for decarbonizing the energy system. Designing possible future scenarios is a common approach to outline the different pathways of how the energy system could be developed, based on varying assumptions about technology and economics. Subsequently, these scenarios shape the energy system analysis conducted by researchers and energy system modelers, influencing the related policy implications. Nevertheless, despite the employment of various methodological approaches, the challenge of inherent subjectivity in scenario-based analysis persists. Most of the studies also fail to engage relevant stakeholders in the process of envisioning net-zero energy systems [1–3].

While efforts are underway to engage the various stakeholders under the umbrella of participatory energy system modeling (ESM), there are little to no studies that engage the vulnerable and marginalized groups, such as young people, who are likely to be impacted most by climate change [3–5]. Moreover, the growing body of participatory ESM literature highlights discrepancies between stakeholders' perceptions and perspectives and those reflected in traditional techno-economic modeling scenarios [6–11]. Although participatory modeling is gaining attention, and policies emerging based on the participatory practices are more likely to be implementable, challenges remain. These challenges include the need to provide information that is both understandable and relevant to participants, design effective stakeholders engagement methodologies, and involve stakeholders from diverse background to ensure comprehensive input. Knowledge gaps among participants can hinder effective communication and understanding, potentially compromising the quality of the insights gathered. Addressing these challenges is crucial to maximise the effectiveness of participatory processes [12]. Additionally, participatory studies that meaningfully engage the public, especially youth—by enhancing climate change knowledge, empowering participants, and collecting socio-techno-economic preferences to facilitate a more inclusive energy transition—remain rare [13]. Furthermore, most participatory modeling studies rely on social science or online survey data to incorporate stakeholders' perspectives. This reliance raises concerns about the validity of the obtained data, primarily due to the lack of prior knowledge necessary to accurately respond to these surveys [14].

Evidence suggests psychological distress about climate change is broadly present among young people, there remains a gap in participatory research regarding the direct inclusion of youth and account their informed choices about green energy transition, i.e., renewable energy technology deployment scenarios. Though participatory research exploring ways to integrate emerging social aspects such as acceptance [15], preference [16], ownership [17,18], and willingness-to-pay [19], the often considered stakeholders are experts from fields like energy/environment [20], health [21], finance [22], tourism [23], education [24], and construction [25]. Moreover, the average number of stakeholders in such studies ranges from 20 to 25, which does not necessarily reflect the opinions of general public [3,20,26]. While the general public featured in less than 20% of reported participatory studies [3], the participants' average age tends to be around forty years [26–28]. Notable exceptions include Holzer et al., who conducted educational workshops with school students (aged 10 to 18) to create Swiss electricity supply scenarios for 2035 [6]. The results revealed that while pupils opted for high shares of renewable energy technologies (RET), their views did not necessarily align with ambitious scenarios proposed by energy experts, who are often the stakeholders reflected in participatory studies. To the best of the author's knowledge, no study has engaged the Norwegian youth in the energy transformation debate to enhance their understanding and assess their choices for future energy systems.

Since 2019, young people have been actively involved in matters related to the green energy transition, as demonstrated by their social media engagement [19,29,30], participation in climate strikes [31], and involvement in national and local climate adoption planning [32,33]. A survey of young people across continents showed that 84% of them are concerned about climate change, and 83% think that people do not take care of the planet [34]. Though young people tend to be less engaged in policy-making, the benefits of their involvement are widely acknowledged. Moreover, youth are generally more progressive than the broader society and actively participate in social movements [35,36]. Studies have also shown that young people are highly receptive to information, and their increased interaction with social media for searching and obtaining detailed information can impact the green transition both positively and negatively [37,38]. Therefore, ensuring their meaningful participation in an inclusive energy transition requires enhancing environmental literacy, and eventually, incorporating their perspectives into energy transition planning [39,40].

To help bridge the gap in exploring youth engagement in energy transition planning, we hold educational workshops with Norwegian high school students (15-16-year-olds). Evidence suggests that information through environmental education can enhance students' knowledge, behavior, and attitudes towards the environment [41]. These workshops allow students to: 1) engage in local, national, and international debates surrounding energy transition, 2) learn about Norway's electricity grid through various interactive activities, and 3) express their preferences regarding social, technical, and economic facets of ESM. The Norwegian electricity system presents an interesting case study for several reasons. Due to increased electrification, Norway's electricity demand is projected to increase by approximately 30% by 2050 [42]. Moreover, the country has the potential to play an essential role in expediting and achieving European net-zero targets, with many studies and reports highlighting its role as a "green battery" for countries like Great Britain, Germany, and the Netherlands [43–45]. However, it remains uncertain whether this potential will be realised, as public discourse in Norway has focused on limiting future electricity exports. Further, Norway is exploring ways to reduce its dependence on oil and gas exports by developing new industries (e.g., battery production), which will require increasing electricity production [46]. Without new production capacities, these new industries will not be developed. It is thus important to understand pupils' views on energy futures in Norway as they will be the ones living in a 2050 decarbonized energy system and will soon be the decision-makers.

While it is possible to engage the youth, caution is necessary as insufficient encouragement to envision the future may provide only a snapshot of their current views. Furthermore, pupils may struggle to provide meaningful inputs without understanding the attributes, constraints, potentials, and environmental implications of RET. Therefore, we divide the workshops into three blocks, beginning with general knowledge about the green energy transformation, climate change, and Norway's current energy system. The complexity increases in subsequent sessions by exploring conflicts of interest related to new RET installations, through interactive activities and an energy justice game.

This study aims to document the engagement process with pupils, and explore how they perceive the information and interactive activities provided during the workshops. Specifically, this work seeks to address two main questions: first, how can young people be effectively engaged in climate change discussions and debates to facilitate a more inclusive energy transition? Second, what characteristics do pupils envisage for a future net-zero energy system or in other words, what possible normative scenarios arise from their perspectives and socio-techno-economic preferences?

The article is structured as follows. Section 2 describes the methods, including a description of workshop materials, the questionnaire, and how the workshops are conducted. In Section 3, the results are presented. Section 4 discusses our findings and summarizes the relevance of using an educational workshop approach to engage youth in a just and inclusive energy transition. Finally, we outline the studys' shortcomings and present our conclusions in Section 5 and Section 6, respectively.

## 2. Methods

This section outlines the methodology for designing and conducting the workshops, detailing the process for gathering information and analyzing the data collected. Our aim was to engage Norwegian school youth (15-16 year-old), introduce them to a structurally designed educational workshop about climate change, green transition, and renewable energy, and understand their perspectives regarding 1) the choice of renewable energy technologies, 2) their spatial allocation, 3) landscape choices for installing RETs, and 4) ESM techno-economic preferences for reaching net zero energy goals. Designing workshop activities is an effective and valuable method for eliciting participants' preferred pathways in the energy transition [47]. Recognizing the importance of details — such as choice of activities, how to conduct, and when to schedule them — we engaged in discussions over the course of one year within our interdisciplinary team and advisory board and sought feedback from social science experts, including those in educational sciences, as well as teachers, high-school and bachelor students. The workshop activities and designed materials are discussed in the following sections.

### 2.1 Workshop description

This study is part of the interdisciplinary research project "Energy for Future", which aims to assess youth perspectives about Norwegian future net-zero energy systems. The participants were upper secondary school students, and the workshops were conducted as a compulsory component of the social science and geography subjects. Studies have indicated that these real-life workshop settings enhance the validity of obtained data, and the learning process can meaningfully impact the attitudes, behaviour, intentions, and knowledge [41,48]. The workshops were conducted during the spring and winter of 2024 in five schools located in Eastern and South-Eastern Norway. A total of 286 students from 12 classes, with class sizes ranging from 12 to 32 students, participated in workshops. Among these participants, 220 students agreed to participate in the modeling questionnaire (discussed below), resulting in a response rate of 77%. The main reasons for some students not responding to the questionnaire were fatigue from workshop activities and desire to take a break, as a break was scheduled immediately after the completion of the modeling questionnaire completion. The Table A1 provides further details about the schools, classes, and participants.

We conducted three blocks of interactive sessions, each lasting 90 minutes, across all schools selected for this study. Three rounds of questionnaires were conducted at intervals during these sessions, with timing determined in the workshop material design phase, to ensure that students had the necessary information to respond effectively. These three blocks were completed within the same day or spread over three weeks, with one weekly information session. Each session's themes, topics covered, and interactive activities are detailed below. The workshops held with the pupils were designed with the following key considerations:

- Presenting relevant information about the energy transition/transformation process, including socio-economic, technological, and global aspects.
- Highlighting the goals of mitigating the climate crisis and involved trade-offs.
- Giving pupils the space to engage with the presented material by asking questions, discussing, presenting their own ideas, giving feedback and challenging assumptions.
- Designing activities for collective learning, such as roleplays, educational games, quizzes, and discussions, to enable informed future energy choices based on a shared understanding.
- Adhering to best practices for engaging stakeholders set out in the literature [4].
- Encouraging youth participation in the climate debate and understanding their attitudes towards energy transition by combining social science with ESM.

The final structure of the executed workshops consisted of following components (Fig. 1):

**Block-1:** This block commenced after administering the pre-questionnaire (social science-focused questionnaire). The session provided general knowledge about the green energy transition, climate change, near-zero carbon technologies, and Norway's current and future energy demand and supply. The aim was to establish a common knowledge base among students.

**Block-2:** This session is built on the information presented in the first block. Students explored conflicts of interest, particularly land-use issues related to the installation of RET.

This was achieved through role-playing a town hall meeting, similar to a model UN setting, where groups of students represented different stakeholders, including the mayor, citizens, environmental organizations, industry representatives, high school students, and citizens employed in the local industry. This interactive activity, centred around the decision on the wind farm installations, allowed students to debate, nuance, and discuss diverse viewpoints on the green energy transition. The block concluded with the administration of the modeling questionnaire.

**Block-3:** This part concentrated on climate justice, including the north-south perspective, and involved a climate-justice game (international climate negotiations) [49]. Students acted as youth representatives of different countries with the goal of achieving international carbon reduction targets, allocating funds from their given budgets for national and international mitigation measures. Students engaged with aspects related to distributive, recognitional, and procedural justice. The workshop concluded with the post-social science-focused questionnaire.

Before conducting the workshops, we conducted a series of four-day pilot sessions with pupils in Oslo to determine the appropriate material and topics to focus on. This process helped us refine the teaching material and choose the activities for the final workshop design based on students' engagement, comprehension of the materials, and learning outcomes.

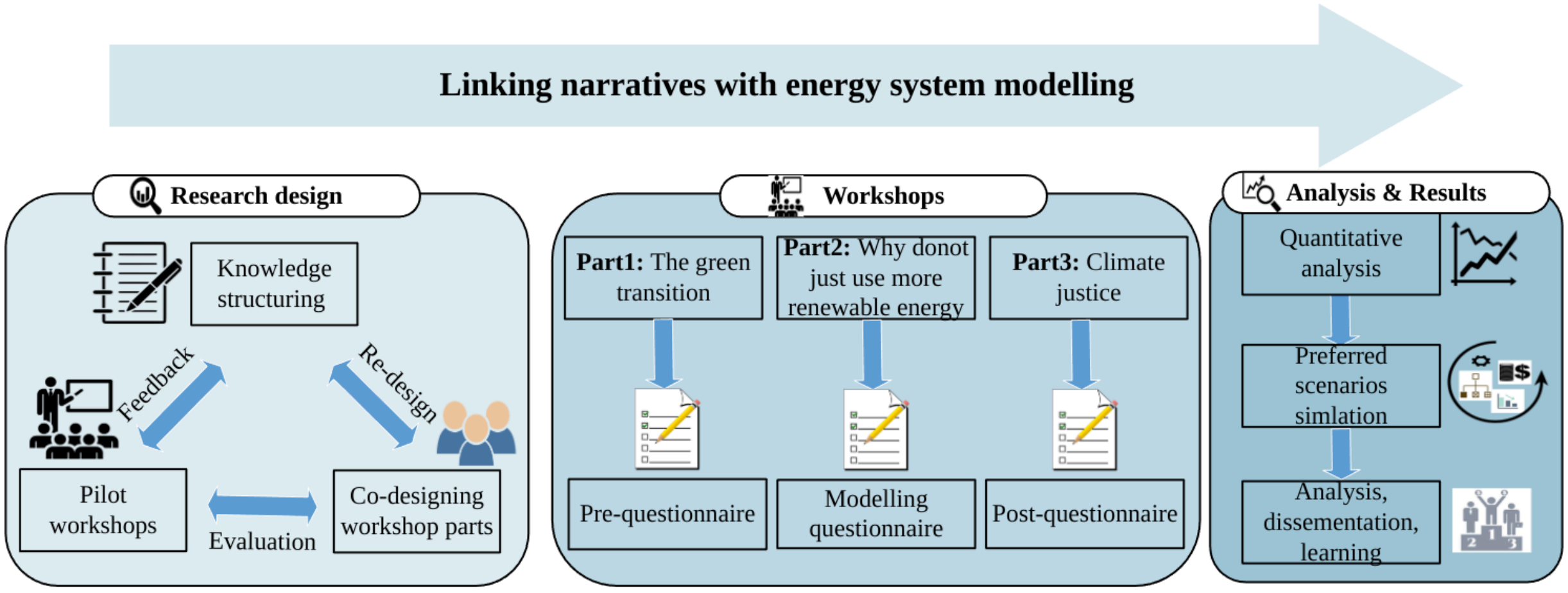

Fig. 1. An optimal participatory approach captures insights from social scientists and energy modelers in designing the knowledge component and the questionnaire/survey segment of the engagement activity. This interdisciplinary study incorporates an iterative feedback process

involving continuous discussions. Qualitative and quantitative insights gathered from these sessions would later be employed with a high spatiotemporal resolution energy system model to capture the trajectories of student preference-based energy systems.

### 2.2 Questionnaire design

Two questionnaires were developed for the students. One was specifically designed to capture students' socio-techno-economic preferences for energy modeling, hereafter referred to as the modeling questionnaire. The other one was a social science-focused questionnaire, measuring students' knowledge, empowerment, and attitudes towards renewable energy and climate change before and after the workshop. In this study, we focus on discussing the design and findings of modeling questionnaire only. The findings related to social-science aspects will be reported in another publication.

The modeling questionnaire was developed in alignment with the structure and quantitative data requirements of an electricity system model that optimises both the investments and dispatch to design a future net-zero electricity system [50,51]. In future work, we aim to simulate the developed scenarios and disseminate insights into how pupils envision the Norwegian net-zero energy system. Considering the substantial time investment required to build energy models, we selected the highRES energy model during the workshop materials design stage [51]. This approach allows us to account not only for model limitations but also for the assumptions necessary to translate the questionnaire output, overcoming the challenge of "institutional inertia" described by McGookin et al. [4]. The modeling questionnaire broadly focused on the specifics of RET: 'what' types of RETs pupils prefer, 'where' they believe these should be installed geographically, 'how' these installations should be managed, and the potential impacts of these relatively unfamiliar technologies on people. The rationale for including socially constructed aspects is supported by several studies, which argue that public engagement should consider not just the physical or technical attributes of a given technology but also the affective and symbolic social aspects, such as visual disruption as well as the feelings and values that people associate with a technology [52–54]. Additionally, ESM modeling attributes such as power generation technology preferences [55], type of power transmission lines [56], and local economy [57] were also considered in the questionnaire.

The modeling questionnaire begins with an open-ended question asking students to identify the most important factors, which, according to them, need to be considered when installing RET. This question helped us gather their initial thoughts, which might be influenced by subsequent information and questions, as noted by Demski et al. in a UK-based public preference study [7]. Demski et al. observed that people's energy preferences could be swayed, when the example scenario or reference point of the scenario-building tool changed. Following the open question, a series of discrete choice questions probe students' technology preferences. Discrete choice is a commonly used interactive method in participatory studies [15]. We employed a variety of question types in our form — including multiple, discrete, single, and scale-based choices — to minimize potential biases in the representation of students' preferences. Fig. 2 illustrates an example of an offshore and onshore choice card. The technology attributes used were selected based on their frequent application in participatory literature (see supplemental file for questionnaire) [12,15,18,32,35,56,64].

Electricity cost-related savings were calculated based on the current average electricity price, yearly average Norwegian household electricity consumption, and wind and solar electricity production costs [60–62]. The visual impact of technologies was considered based on expert views about height and frequency of daily observations, while area calculations were performed using the average electricity consumption of Norwegian households and technology’s area efficiency ($Wp/m^2$). Power transmission lines, a major portion of new solar and wind farms' costs, were considered based on how far the given technology is typically installed from demand centers. Although students discussed wind turbine colors (leading to a conversation about artist jobs), we avoided including such physical attributes (i.e., color and orientation) to maintain a broad representation of socially constructed aspects.

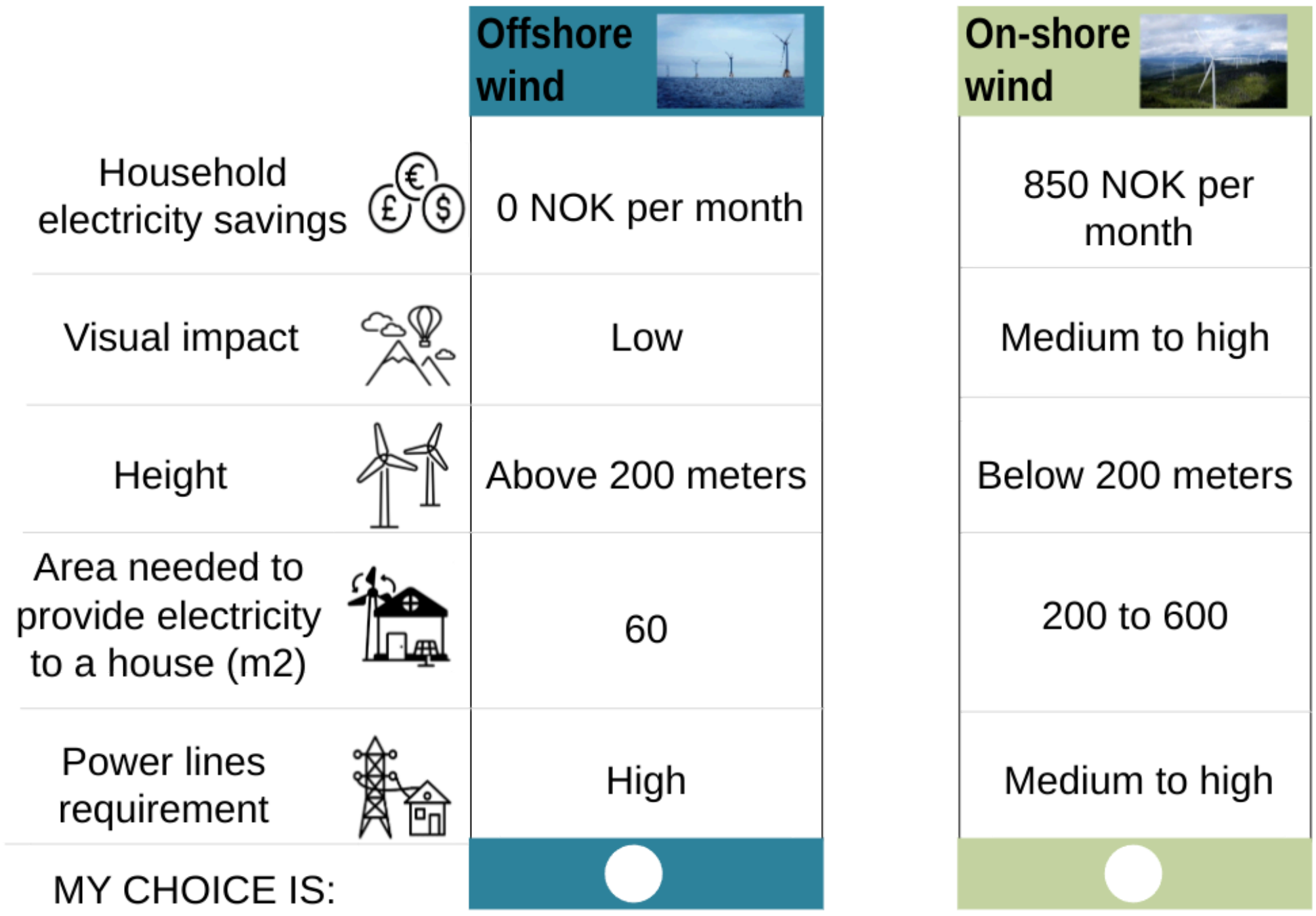

Fig. 2 Exemplar discrete choice card. The full questionnaire is provided in the supplemental file.

Following the technology choices, we added a series of landscape preferences for installing wind turbines, utilizing the Likert scale ranging from 1 to 9, where 1 represents "strongly disagree," and 9 represents "strongly agree." We considered nine landscape types based on the prevalent landscape types in Norway. Using Norwegian national map data and CORINE land cover data, we performed spatial analysis to determine how much these landscapes are distributed across Norwegian counties and in relation to Norways' total area, which we discussed in the results section [63]. Additionally, students were asked to choose their preferred Norwegian counties for solar and wind energy installations by referring to the map of Norway. Subsequently, there were single and multiple-choice questions addressing strategies for overcoming the variability of RET, electricity trading with neighboring countries, and preferred transmission type.

Questionnaire drafts were consistently shared and discussed with social psychology scientists, political scientists, sociologists, RET scientists, and energy system modelers who assisted in shaping the form to minimize technical jargon and make it accessible to school students. This interdisciplinary approach helped to ensure a good fit between the aspects assessed and the energy system model while considering multiple dimensions of the energy transformation process. Brief modeling knowledge and the importance of the questionnaire from a modeling perspective were shared with students before distributing the modeling questionnaire (Block-2). For example, topics such as the cost reduction in RET over time, the land occupied by RET installations, and the consequences of installing RET in remote areas (e.g., high transmission costs) were discussed.

### 2.3 Facilitation of workshops with students

The workshops were facilitated by master students and researchers with a background in social or natural sciences, who were also involved in designing the workshop materials. All workshops' instructors were trained before the activities. The authors of this study also participated in moderating sessions. We discussed with students how energy system modeling outcomes might be visualized on a map of Norway, and how they could be used as guidance for policy-level decisions. The concise teaching/information sessions ensured that students had the knowledge to provide informed and least-biased qualitative and quantitative responses. We noted their observations to build guiding principles for designing and modeling emerging scenarios for use in ESM. Additionally, we explained how the questionnaire results could influence modeling outcomes. Sharing our plan to integrate students' responses into ESM served two key purposes: first, it illustrated how students could contribute to envisioning a future zero-carbon energy system; second, it encouraged them to respond to the questionnaire thoughtfully [64].

Our approach to the workshop design is based on a functional dynamic approach, which allocates different levels of control to stakeholders depending on the objectives [62, 71]. Trutnevyte et al. delineated three modes of information flow during stakeholder involvement: communicating, consulting, and collaborating [72]. In our workshops, teaching sessions served as a mode of communication, while the questionnaire activity fell under the consultation and collaboration category. Additionally, during the collaboration phase, we aim to share model outcomes derived from scenarios shaped by student feedback, fostering a more meaningful integration of stakeholders' preferences into ESM.

## 3. Results

The following sections discuss the pupils' perspectives, socio-techno-economic preferences, and emerging energy scenarios derived based on the data obtained during the executed workshops. One of the key challenges of participatory modeling is the need to make assumptions when designing energy scenarios based on stakeholders' qualitative and quantitative data [65]. Acknowledging these limitations, our aim here is to present students' preferences, which will later be integrated into the spatially explicit electricity system model highRES for Norway to design future electricity systems based on student-driven scenarios [51]. The analysis files are available on GitHub [https://github.com/JavedMS/Mapping_Workshops_EFF], allowing readers to understand the results better.

### 3.1 Technology preferences

Although we did not consider all renewable energy generation technologies, Norwegian youth are divided not only on which technologies to install and to what extent but also on the locations for these installations. As shown in Fig. 3, pupils' choices between location of offshore and onshore wind turbines, with or without the coexistence of solar photovoltaic (PV) panels indicate a strong preference for offshore wind-based solutions. Independent of the solar technology choice, offshore wind emerged as the preferred option to meet Norway's increasing electricity needs in the future. Approximately 35% of pupils selected the combination of solar and offshore wind installations, making it the most preferred choice, while around 33% favored exclusively the offshore wind technology. This indicates that over 68% of the students favored some form of offshore wind as a future electricity source. Given that the Norwegian government aims to install at least 30 GW of offshore wind capacity by 2040, these results suggest that large-scale offshore projects could gain support from youth, given that they are engaged [66]. The discussions with students during interactive activities revealed several reasons for not prioritising onshore wind farms over offshore ones. These include a desire to protect the natural landscapes, minimize environmental impacts, and address visual concerns. For example, during interactive sessions, pupils often discussed issues such as bird collisions, painting wind turbines in different colors to enhance the aesthetic appeal or to create jobs for artists, and avoiding placing wind turbines in forested areas.

Although, approximately 32% of the respondents favored some form of onshore wind-based solutions, there were concerns about installations near residential areas, within forests, and in agricultural areas. These concerns are evident from their landscape preferences and location choices (Fig. 4). We noticed that many pupils' objections stemmed from limited knowledge or misconceptions about RETs. For instance, during an interactive activity, some students discussed issues about the toxicity of wind blade coatings and the belief that wind turbines could only be white. Such concerns mirror the misunderstanding spread by vocal Norwegian anti-wind power movements [67]. While a clear preference for offshore wind emerged, pupils were much more evenly divided regarding the addition of solar power to the energy mix (Fig. 3).

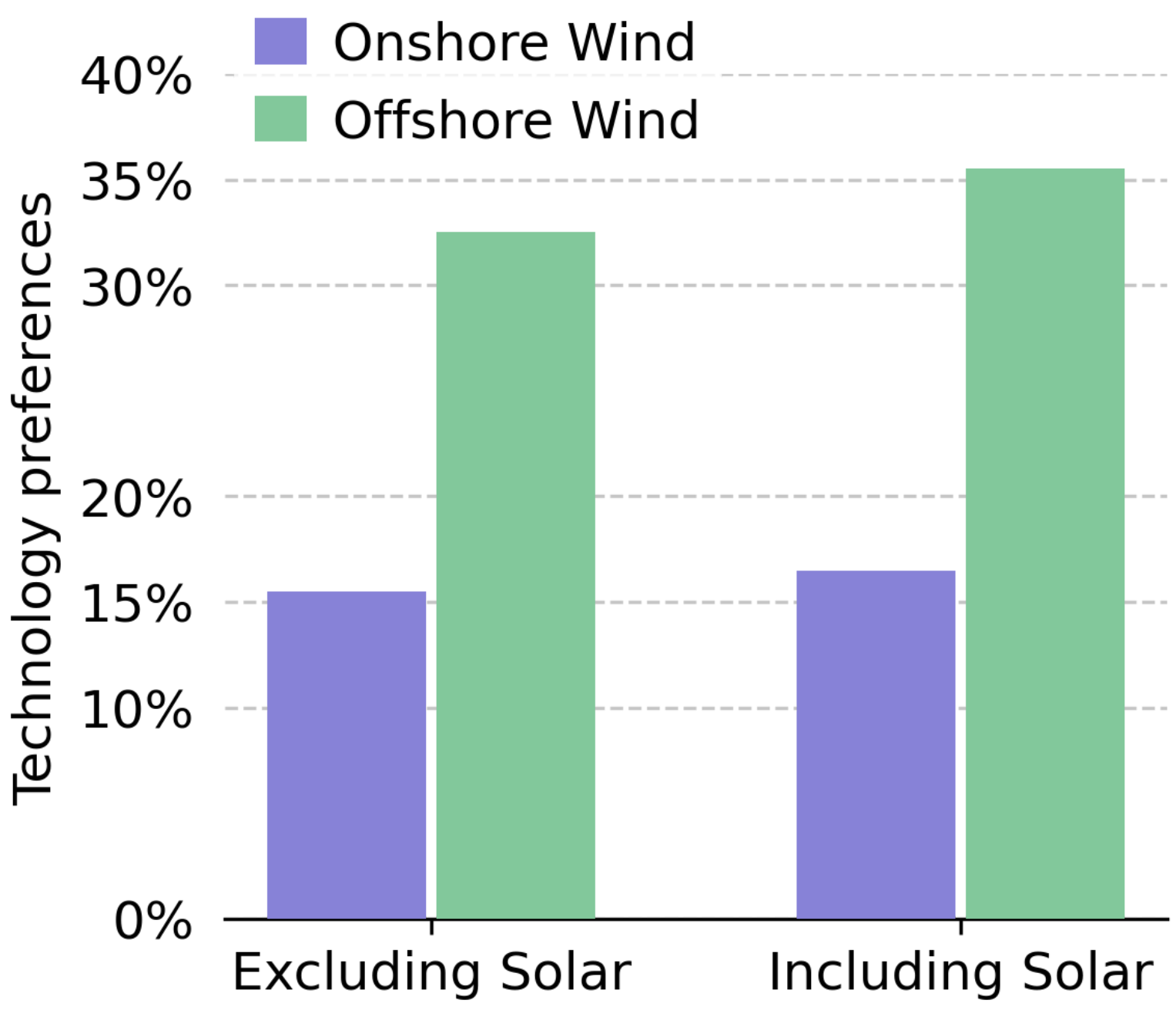

Fig. 3 Representation of pupils' technology choices based on five attributes (savings, visual impact, height/size, required area, power lines [see Methods]), indicating a high preference for offshore wind, with or without solar.

Pupils have the potential to raise awareness among the general public about the characteristics of different technologies and their potential impacts, leveraging their active use of social media [19]. Rejection of wind energy in local municipalities have recently been observed in Norway. Youth from these areas could clarify doubts and provide updated information if the group is meaningfully engaged in decision-making processes about new RET installations.

### 3.2 Spatial and regional technology preferences

Fig. 4 provides an overview of pupils' landscape preferences, while Fig. 5 details the spatial distribution of their preferences concerning landscapes and regional choices for the installation of solar and wind farms across Norway. The results indicate a strong preference (median > 6) for installing wind turbines in the sea area as well as mountainous and vegetated regions. Pupils also showed moderate agreement (median = 5) for placing wind turbines in natural grasslands, nearshore locations, and industrial areas. Conversely, there was apparent disagreement (median < 4) regarding installing wind farms in forests, agricultural lands, and residential areas. Beyond legal obligations, Norwegian culture and politics may influence pupils' inclination to conserve natural landscapes, such as forests, with a focus on prioritizing biodiversity and nature preservation for future generations. However, research indicates that only 1.7% of Norwegian forests remain untouched by forestry activities, with the remainder having undergone at least one cycle of clear-cutting [68]. This process involves systematically removing all or most trees from an area to enable forest regeneration or the establishment of new plantations. This underscores the need to effectively communicate the current state and ecological role of these landscapes in biodiversity and nature preservation to the public, aligning perceptions with the actual state of landscapes and potentially reducing opposition to the installation of the land-based wind turbines.

As depicted in Fig. 5, there is a higher tendency among pupils to install more onshore wind in the Northern, Central, and Western regions of Norway, compared to the Southern and Southeastern areas, where solar technology is preferred (Fig. A1). It is noteworthy that Norway's most suitable wind energy sites, including existing wind farms, are also located in central and western Norway. Similarly, most of Norway's solar potential is concentrated in the Southeast and East, aligning with the pupils' preferences for solar installations. Given that over 55% of Norway's population resides in the Southern and Eastern parts of the country, significant solar energy potential could be harnessed through rooftop-installed PV panels. In the future, pupils could be a key in promoting acceptance for this RET by representing their communities in local municipalities. These results underscore the importance of engaging local pupils in policy-shaping, ensuring they are not left behind. For instance, 93% of pupils, who selected Oslo expressed a preference for solar PV installations, while 70% of those, who choose Nordland (Northern Norway) believed that wind turbine installations would be more suitable for this county (Fig. 5). However, as the workshops were conducted in schools in Eastern and South-Eastern Norway, these results cannot be generalized to reflect students' preferences nationwide.

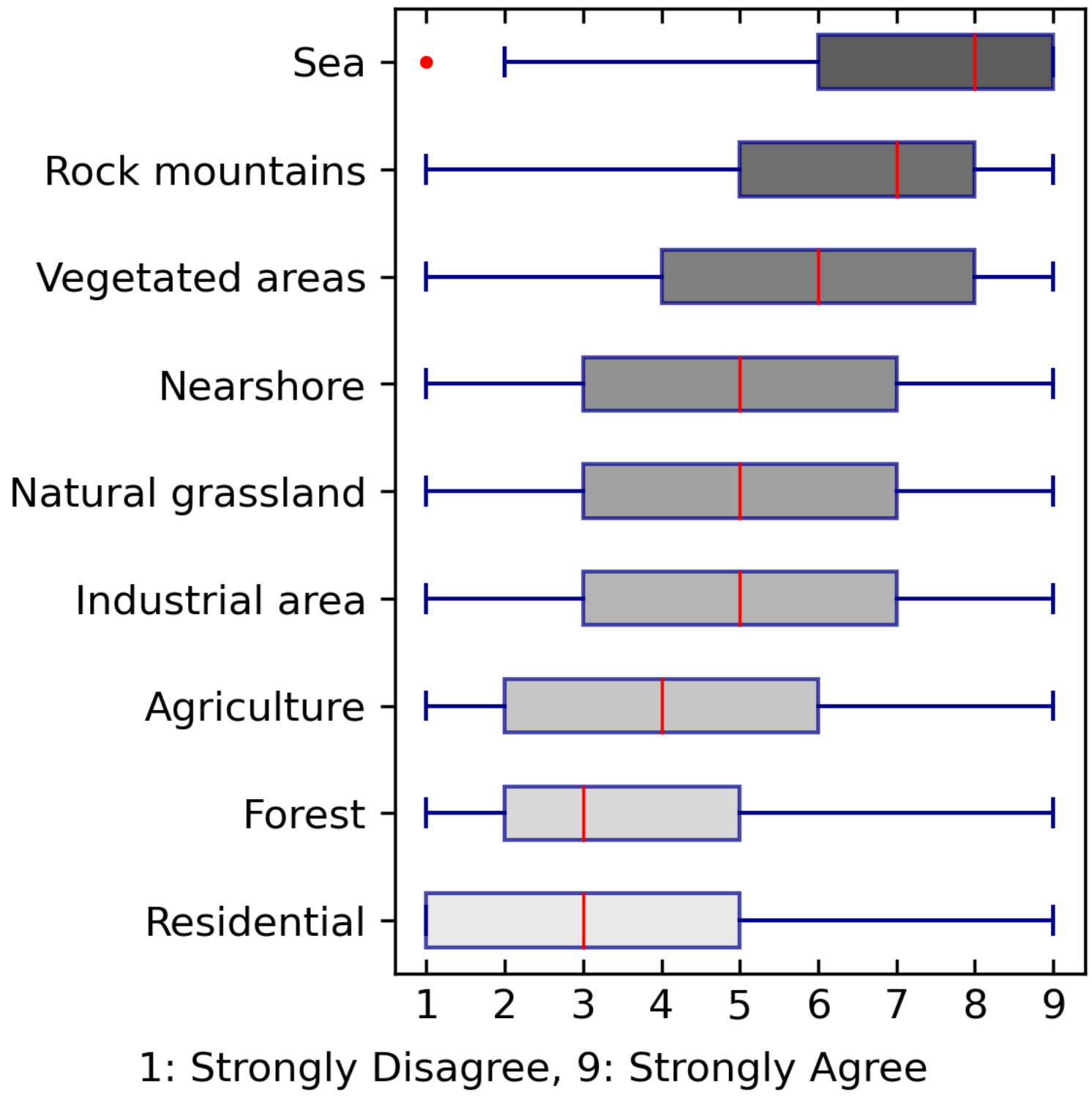

Fig. 4 Pupils' agreement with placing wind turbines in different landscape types, rated from 1 (strongly disagree) to 9 (strongly agree). The boxes represent the distribution of responses,

with the central read line indicating the mean, while the whiskers illustrate the range and variability of responses. Pupils' landscape preferences regarding wind-based technology reveal that, on average, they are positive towards placing wind turbines in the sea and on rocky mountains, but they disagree with having them in forests or near residential areas. Opinions are more divided regarding the other landscape types, with higher variance and a mean towards the scale's midpoint.

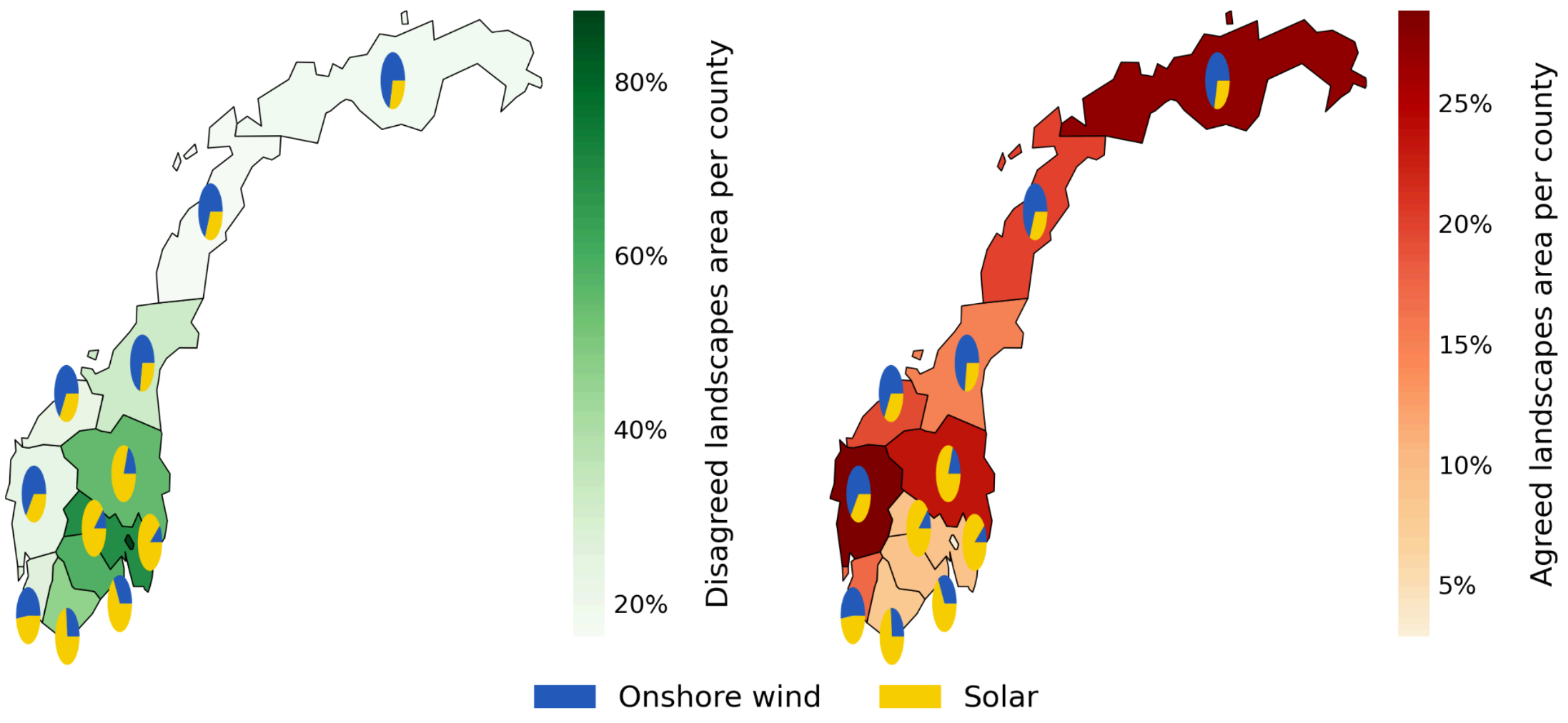

Fig. 5 Spatial distribution of pupils' landscape and RET preferences. Fig. (a) shows the percentage of each county's area with unpreferred landscapes (see Fig. 4), while Fig. (b) illustrates the percentage with preferred landscapes. Pupils preferred wind energy installations in counties with a higher concentration of preferred landscapes, whereas PV technology in densely populated areas with the predominance of unpreferred landscapes (i.e., south and southeast Norway). See also Fig A1.

The spatial distribution of pupils' technology preferences aligns with the likely location of wind and solar farm in Norway. In general, pupils' responses to statements about area preferences align with their responses to map-based questions about landscapes and counties. However, some differences are noteworthy (Fig. 4, Fig.5, Fig. 6). Fig. 6 shows that for each statement-based question, at least 33% of respondents neither agree nor disagree with placing wind turbines in the specified areas. For instance, ~40% neither favored nor disfavoured the placing of wind turbines in remote areas. Yet, when it came to map-based location choice questions (Fig. A1), Northern and Central Norway counties (with only ~16% of the Norwegian population) were the most selected for the installation of onshore wind turbines (i.e., remote areas). Western Norway emerged as pupils' second most popular choice for the onshore wind technology. The difficulty to make a decision reflected in the 'neither agree or disagree' response might stem from using various terminologies, such as 'remote areas' in statement questions compared to county names in location-based questions. This might lead pupils to consider the vast availability of land in Northern Norway or to associate the specific county names known for wind turbine placements or specific counties known for good wind conditions. Similarly, ~43% of students were uncertain about 'areas where people' in statement-based questions, yet they clearly expressed their opinions in map-based landscape questions (i.e., industrial area and residential area), especially strongly opposing the presence of wind turbines in residential areas.

Despite significant efforts involving interactive activities, discussions, and systematically designed questionnaires, this high percentage of indecisive responses may arise from various underlying factors. Firstly, pupils might have found it irrelevant to respond to the same topic twice and therefore choose neither agree nor disagree, as they have responded to it in picture-based landscape and map-based location preference questions. Secondly, viewing images tends to be more engaging, potentially leading to more immediate and informed responses regarding landscape preferences [69]. In contrast, textual statements rely on individual imagination and interpretation, which can vary widely and may result in varied understandings of the same question. Finally, while a lack of question understanding or relevant knowledge might also have contributed to answers obtained, it was not the case for everyone, as at least more than 57% of pupils responded to each statement-based questions with either agreeing or disagreeing (Fig. 6). While most pupils selected neither agree nor disagree in statement-based area preference questions, those who agreed or disagreed exhibited preferences that aligned well with their landscape, location, and technology preferences. For instance, pupils agreed with the idea of not placing wind turbines in nature reserves and people living areas, which correlates with their opposition to the utilization of the residential and forest landscapes (Fig. 5, Fig. 6). Similarly, the sea, as the most agreed-upon location, aligns well with their strong offshore wind turbine preference.

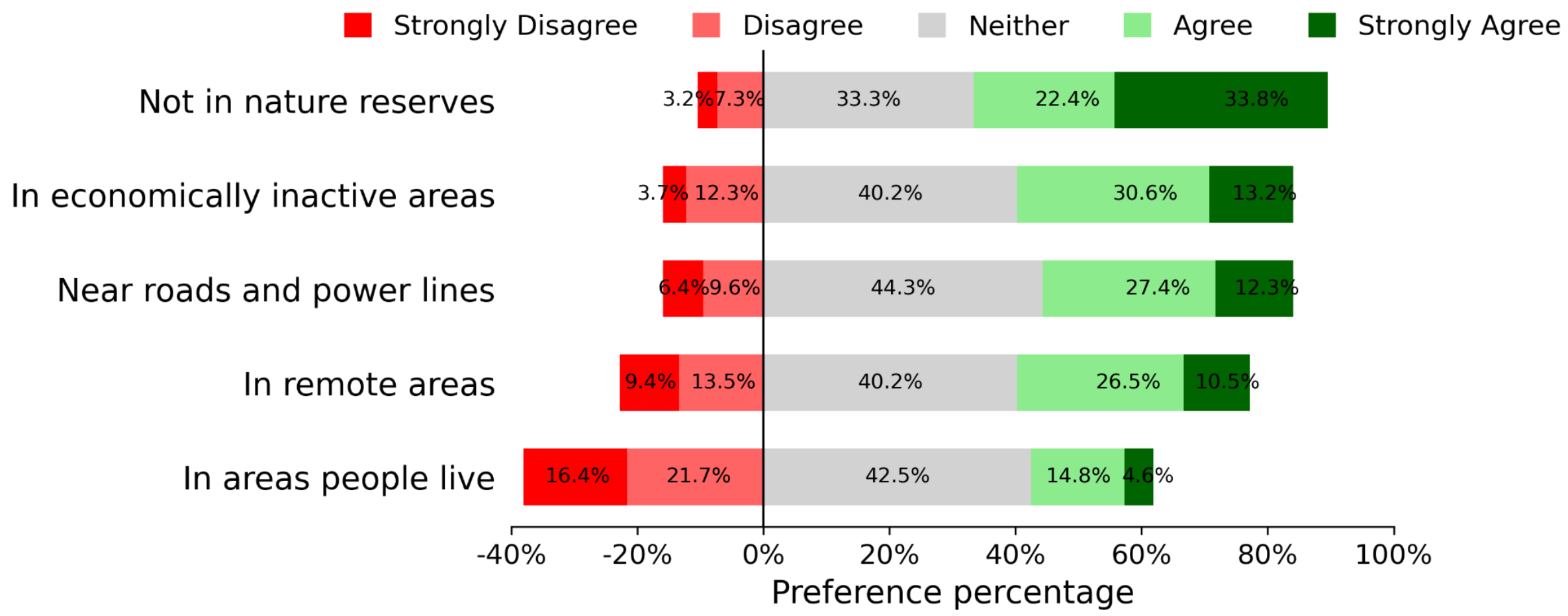

**Fig. 6** Pupils' choices about placing wind turbines in different area types, ranging from existing infrastructure and developed areas to remote and nature reserve areas. Most pupils choose neither agree nor disagree, illustrating either pupils are indeed unsure or lack interest. Abstract terms are used in this figure to summarize the statements used in the questionnaire.

### 3.3 Techno-economic choices

**Fig. 7** depicts the pupils' preferences regarding techno-economic parameters often only assumed in ESM. Pupils appear reluctant to adopt a demand-side management approach to mitigate the variability of RET, as it may directly impact their lifestyle. Instead, they preferred more expensive options such as installations of renewable energy storage solutions and electricity trading with neighboring countries. The International Energy Agency has projected that 500 GW of power demand response capacity will be brought to the markets by 2030. However, it's crucial to proceed cautiously, as people may hesitate to change their electricity consumption patterns unless they are fully engaged and adequately compensated [70]. Additionally, pupils aim to ensure energy independence (i.e., self-sufficiency) with preferences for electricity trading and energy storage. Approximately 42% favored a balanced approach to electricity import and export, suggesting that imports should be equal to exports. This was followed by 32% and 26% saying "Import same as today" and "Import more than today". These preferences align with the open-ended question responses (Fig. 8b), where terms such as "environment", "nature", and "protection" were frequently used, highlighting that sustainability is the most preferred priority. Regarding power transmission lines, pupils were divided between overhead and underground options. Despite this division, there is minimal opposition (<4%) to installing new power lines.

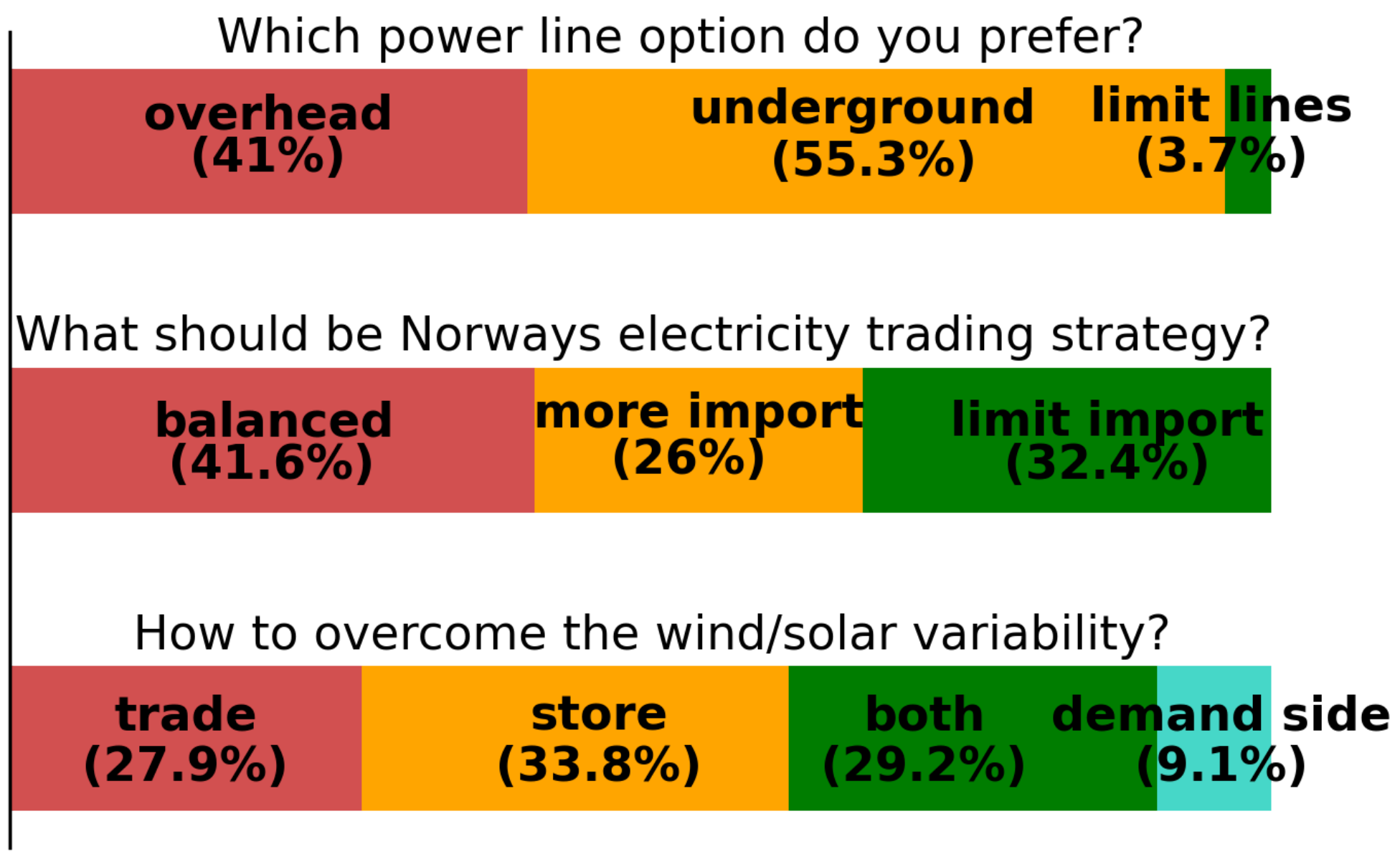

**Fig. 7** Pupils' choices for technical parameters are often considered energy model assumptions. Pupils are aware of renewable energy's intermittency and prefer its storage/trading while ensuring energy independence (middle bar). Additionally, there is no strong opposition to installing new transmission power lines.

**Fig. 8a** illustrates pupils' choices concerning the important factors in electricity production. It is evident that self-sufficiency, or energy independence, is their top priority, followed by zero-carbon emissions and environmental protection. Fig. 8b complements these priorities by presenting pupils' responses to an open-ended question about the three most important factors to be considered when installing renewable energy technologies. All these factors are essential parts of ESMs. ESMs typically aim to minimize total energy system costs while having several constraints, such as emission reductions, trading with other countries (self-sufficiency), and limiting the capacity potential of RET due to, e.g., biodiversity protection. However, interestingly, cost-minimization, which is at the core of most models, is deemed the least important factor by young people. To ensure an inclusive energy transition, energy models need to also consider public preferences across all model decision variables, implementing constrained or multi-objective optimization. For instance, regardless of solar/wind capacity factors, locations concerning biodiversity can be excluded with high spatial resolution to optimize viable locations. Similarly, preferences for areas can be modeled by constraining the model to prioritize preferred locations before considering others.

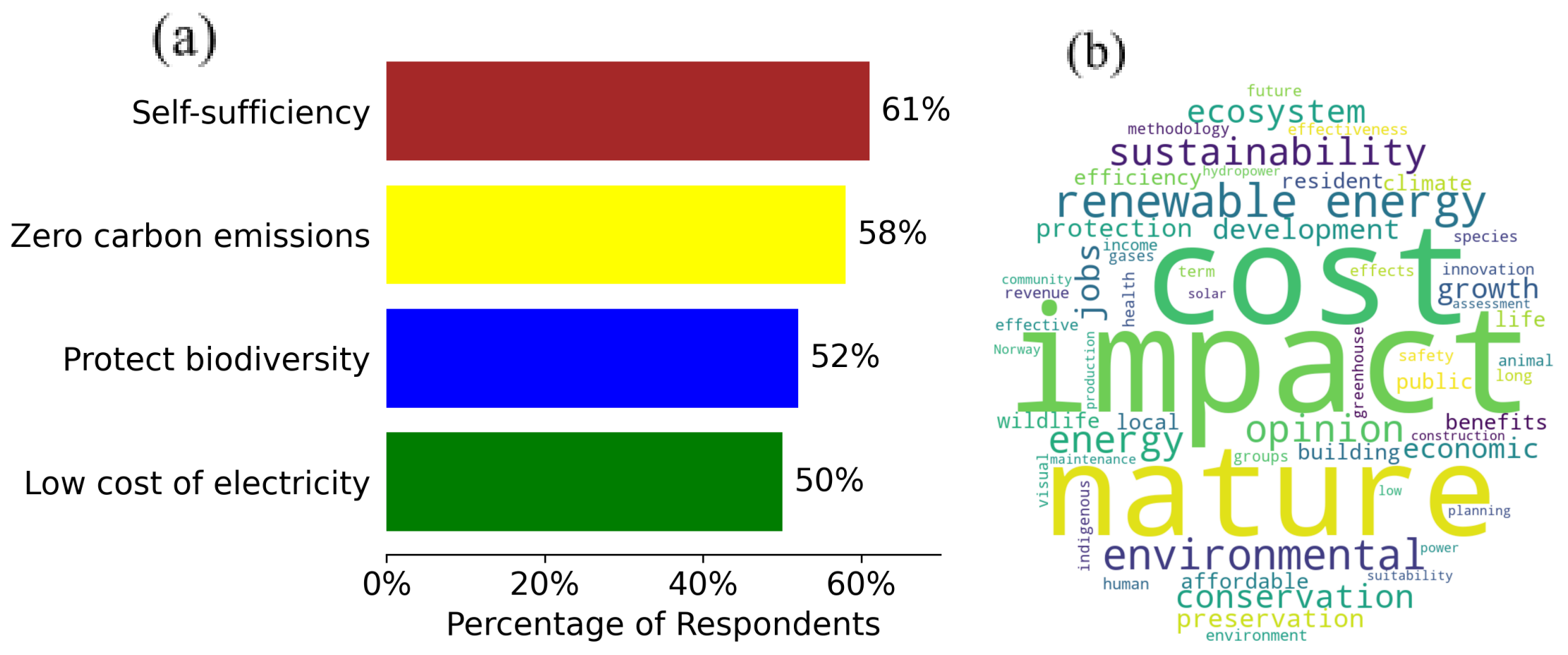

**Fig. 8** (a) Energy models that often prioritize the least-cost objective for optimization are labeled the least prioritized by pupils. (b) The open-question results reflect their primary concerns, which are nature and economy.

## 4. Discussions

Although participatory research in ESM has gained researchers' attention, McGookin et al. observe that "*it is a poorly developed practice...*" and emphasize adhering to good participatory practices [3,71]. In this study, We followed the experiential guidance for engaging stakeholders and collecting qualitative and quantitative data as outlined in the literature, which is based on the insights from experts in the fields of ESM and participatory research [4]. Table 1 below highlights how these practices are perceived in this study.

Table 1. Good transdisciplinary practices were adopted for conducting workshops [4].

| Practice | Explanation | Action |
|---|---|---|
| Stakeholder mapping | Who engaged?<br>Why engaged?<br>What is their role? | Young school students<br>Informed energy transition preferences<br>Non-expert stakeholder |
| Flexible/adaptive approach | Respond to stakeholder needs. | Conducted pilot school workshops to refine the questionnaires. |
| Acknowledge model limitation | What model can and cannot model/optimize/predict?<br>What social aspects can be modeled?<br>How and which type of constraints can be implemented? | highRES model capabilities and limitations were considered during the workshop material design. |
| Inclusion of divergent views | Respect conflicting views | Debate and roleplay based on a real case in workshops |
| Be critical of the whole process | Avoid box-ticking exercise | Both transdisciplinary and interdisciplinary approaches are used, engaging in several rounds of constructive feedback. The pupils were aware of their role in the research process |

To engage the youth in green transition debate, this study illustrates the preferences of teenagers in Norway. The results obtained in this work demonstrate that pupils showed a preference for offshore wind over onshore wind technology, with approximately 68% of them choosing offshore wind parks, with or without additional solar technology installations (Fig. 3). The pupils' strong preference for offshore wind aligns with recent studies in Norway, where people preferred nearshore and offshore wind installations, given that ownership of the projects should remain at local and national levels [18]. This could support the Norwegian government's 30by40 offshore wind development goal of hosting 30 GW offshore wind power capacity by 2040 [72]. However, the potential challenge of offshore areas would be public opposition due to their use for commercial fishing activities and the presence of ecological valuables such as spawning grounds, nearby bird nesting sites, and protected areas, as well as high costs [66].

Pupils' preference for offshore wind can be understood in the context of protests by Norwegian people since April 2019, when the Norwegian Water Resources and Energy Directorate (NVE) proposed a map indicating the most suitable onshore wind park locations amounting to 29,000 km2, or roughly 9% of total Norway land area, corresponding to approximately 290 GW of the onshore wind generation capacity [73]. The proposed plan was scrapped due to the strong opposition from environmental associations, ministries, and municipalities. Additional reasons for favouring offshore wind may include knowledge of favourable sea wind conditions and concerns about noise and visual impact. Further studies across national levels are essential to gain a deeper understanding of the underlying motivations for wind energy preferences.

Generally, international opinion polls highlight positive public perceptions about renewable energy. Still, at the local levels, people oppose different technologies across various social and cultural contexts [52]. The workshops with students reflected their choices about technologies for given municipalities. To avoid biased opinions, we didn't specifically present country-wise solar or wind energy availability (i.e., capacity factors) in Norway, but we discussed the importance of wind speed or sun availability with students during sessions. By coinciding technology choices with location and landscape choices, some patterns emerge, such as strong disagreement with installing wind turbines in forests and residential areas, concentrated mostly in the south and east of the country. The result that students preferred wind power in central and northern Norway might be due to the students coming from southern Norway, reflecting some NIMBYism. However, it is important to note that recent studies have shown that NIMBYism is overly simplistic, and public responses about RETs are much more complex [52,74]. Alluding to selfishness may miss important insights for no acceptance at the local level. The pupil spatial choices are interesting given that the workshops were held just a few months after the agreement between the government and the Sámi people in the Fosen case was widely discussed in Norwegian media. This case highlighted the area conflicts over central and Northern Norway between reindeer herding Sámi people and wind installation development.

Pupils expressed eagerness to have wind turbines installed in areas with little economic activity to boost local business growth (Fig. 6). However, results from an interactive activity asking about localized consequences of the green energy transition revealed concerns about potential reduced economic growth. These seemingly conflicting perspectives show that while the young generation wants to support the growth of local economies by installing new wind farms, they also fear local economic stagnation due to these installations. This concern may stem from perceptions that the economies based on RETs might not generate a significant number of local jobs or that the presence of large-scale infrastructure could detract from other types of economic activities. For instance, a common discussion during workshops was whether RET installations would result in local job creation, with concerns that they might not create new opportunities. This suggests students foresee that place-based economies may not necessarily be part of a sustainable transition. Discussions with students indicated that while climate change is a realistic concern, solutions should not come at the cost of destroying nature or local economies.

A valuable outcome from this exchange with pupils is an insight into their understanding, thoughts, views, and apparent contradictions in their choices and visions about the energy transformation process. For instance, most students think nature preservation is important, but they desire a self-sufficient, zero-carbon, and cost-effective electricity system in Norway. The landscape choices highlight this apparent contradiction, as most types received below-average or average consideration for wind turbine installations (see Fig. 4). This leads to two recurring challenges. Firstly, landscape types and counties selected by pupils for wind/solar technology installations do not necessarily have high wind/solar capacity factors, requiring more capacity installations to meet the same energy demand. Secondly, such less efficient RET installations would require high energy storage capacities or increased interconnection with neighboring countries, potentially compromising the energy independence criteria.

This divergence in pupil choices highlights an area for energy modelers to explore trade-offs and address uncertainty, ensuring a socially just transition. This can be achieved through various modeling techniques discussed in the literature, which involve coupling spatial, technological, and economic modeling elements with stakeholders' choices [75,76]. This may help assess the cost implications of an inclusive energy transformation and address stakeholders' preferences, such as who would be the primary users of locally generated wind energy—either national or international. In future work, we will simulate the scenarios emerging from pupils' choices to design Norways' future electricity system using highRES electricity system model [51]. To ensure meaningful stakeholders' engagement, we aim to gather students' feedback on the model results, and assess the likelihood that pupils might reconsider their RET and landscape preferences based on the modeling outcomes. The example scenarios emerging from the workshops are illustrated in Table A2.

Having seen the increased discussion in social media about wind energy in Norway, pupils can contribute to understanding the social dilemmas related to wind energy and local opposition to wind technology installations [19]. Pupils' participation in projecting future net-zero energy systems may not guarantee its public acceptance. Still, it can empower the young generation and increase not only their trust in the transformation process but also the ownership of decisions. Psychological ownership increases public willingness to accept wind energy impacts on the local communities, as Dugstad et al. found that people's willingness to accept wind energy depends on their degree of psychological ownership [17]. The ownership of RET plants could be defined as who maintains the control or uses the produced electricity [17,18]. The ownership in our study can be interpreted as self-sufficiency (i.e., the intended user), where 76% pupils either favor a balanced approach between the import and export of electricity or no change in the current status (Fig. 8a). This preference for limiting net electricity imports aligns with the findings of Xexakis et al., who identified a similar consensus among young people from various European countries [77]. The observed pupils' reaction towards RETs can be linked to the pupils' association that comes into their minds when they think about wind power. During the workshops, the response to the question “What comes into their mind when they hear about wind turbines?” was mostly negative, i.e., noise, nature, big, etc. Decision-making theories suggest that complex interactions exist between the cognitive and affective components of the human brain [69]. Pupils likely used affective reactions rather than cognition ones about onshore wind, as discussed by Holzer et al. in the study of making scenarios with pupils about waste incineration [6].

Interactive discussions with students and analysis of their questionnaire responses reveal that while concerns about climate change are valid, efforts to address it should not compromise nature, biodiversity, local economies, or people's trust. The situation calls for a broader understanding of the sustainable energy transformation that addresses all economic, social, and environmental issues. The presented results could help us understand how to align climate actions well with the views of young people who would face the outcomes of current policy planning. Although results highlight the importance of engaged decision-making for fair and just changes, it is and will remain a challenge to effectively bring young generations' diverse and contradictory views into policy making.

## 5. Shortcomings

This study has several limitations that will inform our future research. First, responses to the questionnaire can be swayed by how information and questions are conveyed and formulated. Despite interdisciplinary efforts to minimize bias in the workshop materials, future work may involve designing a more comprehensive educational process where students may revisit concepts multiple times and deepen their understanding of the risks and benefits related to each technology implementation before expressing their preferences. Ideally, this would be based on workshops that extend over several days; however, the student's curriculum constraints may limit the time available for such activities. Second, their existing geographical knowledge might influence pupils' responses about selecting landscapes and counties for wind or solar technology installation. This could lead to biased selections based on known locations rather than objective assessments. Although participants were instructed to respond based on their understanding and preferences rather than biases, we recognize that this is challenging to achieve.

Third, limited time restricted our ability to address all aspects of a socially just energy transition raised by experts or pupils during workshop discussions. For example, there is no straightforward answer to questions like "Does Norway's electricity price increase when electricity is sold to neighboring countries?" as this depends on various factors, such as energy market prices and demand. Finally, our study focused solely on solar and wind technologies. This focus was due to the Norwegian government's plans for wind and solar power, their potentially lower costs, and limited time to explore other technologies, such as nuclear and geothermal, for achieving net-zero targets. This focus was adequate for our study, which aimed to document Norwegian pupils' preferences for the first time and to broadly understand their perspectives and socio-techno-economic choices, often part of energy experts' assumptions.

## 6. Conclusions

Despite the active participation of youth in climate change issues and their vested interests, young people are often overlooked in participatory processes. To address this gap, we engaged Norwegian high school students, revealing that young participants are not only interested in energy-related and environmental topics but are also capable of meaningful engagement in the participatory process.

We demonstrated that youth prioritize energy independence and environmental preservation, such as biodiversity, over energy system costs—traditionally central to energy system models. While about 15% of respondents favored exclusively onshore wind installations, there was strong apprehension about the location of wind farms near agricultural, forested, and residential areas. Independent of solar technology, offshore wind emerged as the preferred option to meet Norway's increasing electricity needs, with almost 68% opting for it. Although pupil's preferences were divided between overhead and underground power lines, there was minimal opposition (<4%) to installing new ones.

To achieve a more inclusive and equitable transition, modelers and decision-makers should incorporate scenarios based on pupils' socio-techno-economic preferences. This study provides an educational workshop-based framework to bring youth perspectives into general debates and accounting them for optimizing and simulating future energy systems,which in turn, will strengthen the net-zero projections. Policy-makers can utilize the presented workshop-based approach to engage youth during school hours to understand and reflect on their perspectives on the evolving energy systems. The benefits of these participatory practices with youth are twofold: firstly, these meaningful engagements offer an opportunity to address young people's growing distrust and help build trust, and secondly, repeatedly engaging youth can cultivate a mutual understanding of the challenges and foster shared ownership of decisions across all aspects of the energy system, thereby promoting inclusive and swift energy transition.

## CRediT authorship contribution statement

Muhammad Shahzad Javed: conceptualisation, methodology, data curation, analysis, visualisation, writing—original draft; Karin Fossheim: conceptualisation, methodology, data curation, writing—review & editing; Paola Velasco-Herrejón: conceptualisation, methodology, writing—review & editing; Nikolai Elias Koop: data curation, analysis; Matylda N. Guzik: conceptualisation, methodology, writing—review & editing, funding acquisition; Charles Dana Samuelson: conceptualisation, methodology; Beate Seibt: conceptualisation, methodology, data curation, writing—review & editing, project administration, funding acquisition; Marianne Zeyringer: conceptualisation, methodology, data curation, visualisation, writing—review & editing, supervision, project administration, funding acquisition.

**Declaration of competing interest**

The authors declare that they have no known competing financial interests or personal relationships that could have appeared to influence the work reported in this paper.

## Supplementary data

The complete modeling questionnaire is provided in the supplemental file. The code used to analyze and plot the workshop data is available on GitHub.

[https://github.com/JavedMS/Mapping_Workshops_EFF].

## Data availability

Workshop materials, including presentations and interactive activities, can be provided on request.

## Acknowledgments

This study is part of the Energy for Future project, funded by UiO: Energy & Environment. The funding institution was not involved in conducting the workshops. We thank all the schools for organizing them and the students for their contributions. Additionally, we extend our gratitude to the workshop instructors who facilitated the workshops and to all individuals who contributed to designing the workshop materials and providing support throughout the project.

## Appendix

Table A1. Students and schools overview.

| School | Number of classes | Number of participants | Workshop format | Location | Study program |
|---|---|---|---|---|---|
| 1 | 1 | 21 | Spread out | Urban area | General |
| 2 | 1 | 12 | Thematic day | Rural area | Vocational |
| 3 | 3 | 60 | Thematic day | Rural area | General |
| 4 | 1 | 31 | Thematic day | Urban area | General |
| 5 | 6 | 162 | Thematic day | Rural area | General |

Table A2. Representative scenarios of pupils' choice.

| | Total technology capacity (percent of new capacity additions) | | | Landscape preference | Location-based technology preference (Total=11 counties) | | Power lines | | Import /export |
|---|---|---|---|---|---|---|---|---|---|
| | Solar | Onshore | Offshore | | Solar | Wind | OH | UG | |
| Base scenario | OPT | OPT | OPT | OPT | OPT | OPT | OPT | OPT | OPT |
| Scenario #01 | 25.8 % | 23.5% | 50.7% | OPT | OPT | OPT | OPT | OPT | No change |
| Scenario #02 | 25.8 % | 23.5% | 50.7% | Exclude disagreed | OPT | OPT | OPT | OPT | No change |
| Scenario #03 | 25.8 % | 23.5% | 50.7% | Preferred landscapes | OPT | OPT | OPT | OPT | No change |
| Scenario #04 | 25.8 % | 23.5% | 50.7% | Preferred landscapes | Pupil choice | Pupil choice | OPT | OPT | No change |
| Scenario #05 | 25.8 % | 23.5% | 50.7% | Preferred landscapes | Pupil choice | Pupil choice | 41% | 55.3% | No change |
| … | … | … | … | … | … | … | … | … | … |

**Technology preference:** The given technology preference is translated as a percentage of total new capacities, set as the upper bound for the specified technology.
**County preference:** The same principle applies to location preferences, where the technology preference for a particular county is set as an upper cap. For instance, if the model opts to include onshore wind capacity in Oslo, it should not exceed 10% [Fig 5].
**Power lines:** As pupils seem divided between overhead or underground options, new transmission types and capacities will be optimized accordingly.

OPT: optimal; OH: overhead; UG: underground; No change: same as today

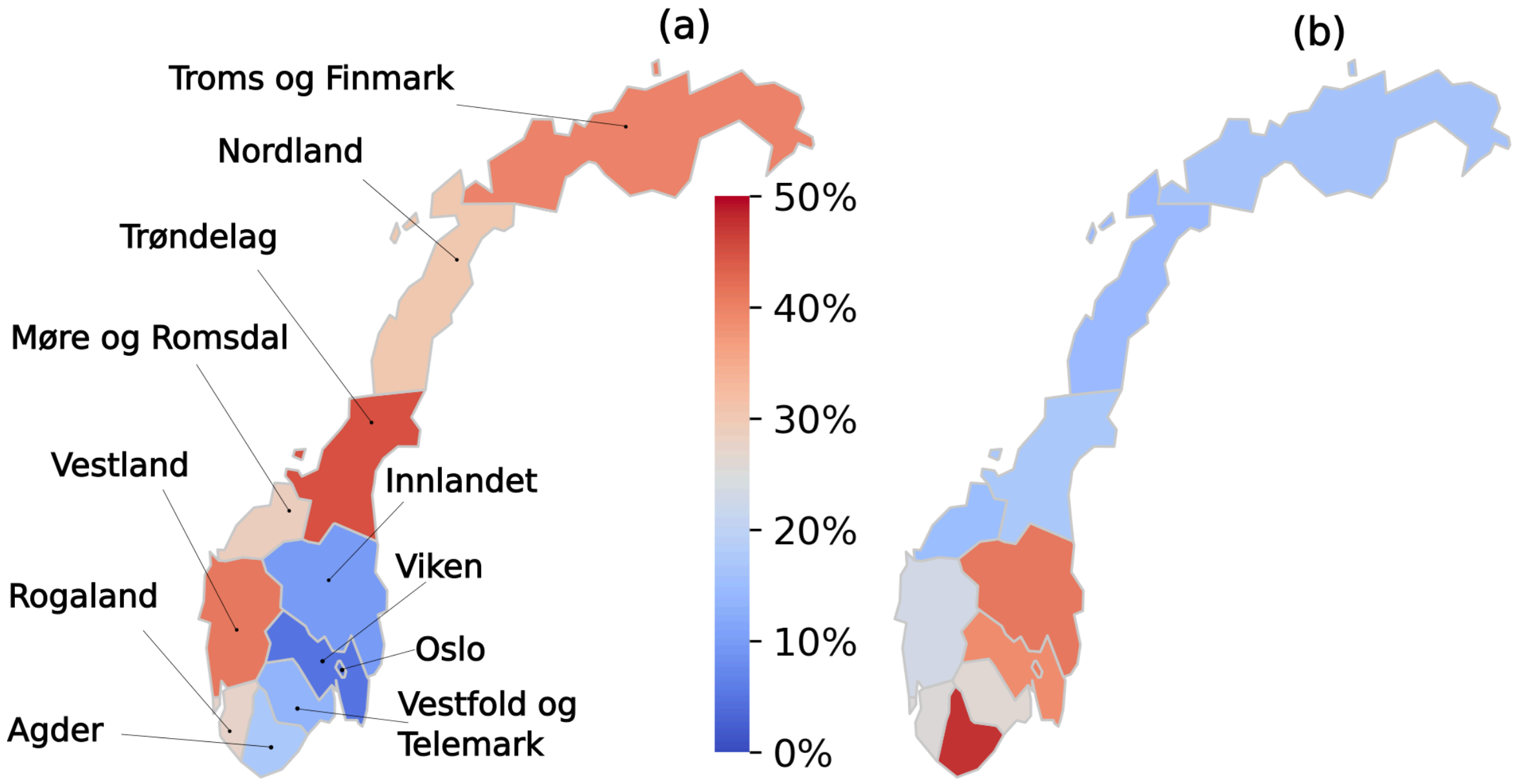

Fig. A1 Pupils' regional preferences for (a) wind (b) solar installations across Norway. Nearly 50% of participants selected Agder county for solar installations, while approximately the same percentage preferred Trøndelag county for onshore wind installations (Related to Fig. 5).

# Beyond costs: Mapping Norwegian youth perspectives for inclusive energy transition

## Modelling Questionnaire

# Supplemental file

Muhammad Shahzad Javed[1,*], Karin Fossheim[2], Paola Velasco-Herrejón[1], Nikolai Elias Koop[3], Matylda N. Guzik[1], Charles Dana Samuelson[2], Beate Seibt[2], Marianne Zeyringer[1]

[1]Department of Technology Systems, University of Oslo, Norway.

[2]Department of Psychology, University of Oslo, Oslo, Norway.

[3]Department of Mathematics, University of Oslo, Oslo, Norway.

[*]Corresponding author(s). E-mail(s): m.s.javed@its.uio.no

**Question for all:**

According to you, what is the most important to you that should be considered when installing renewable energy technologies? Rank them 1 as the most important and 3 as the least important (i.e., cost, nature reserves).

1

2

3

**Table attributes explanation**

**Household electricity savings:** Spending less on home energy because wind and solar power are cheaper now. This can also lead to more local municipality benefits, i.e., sports facilities of people's choice.

**Visual impact:** How often the installed wind/solar farms will be exposed/seen by local people. There are three categories of visual impact: High (seen by many people); Medium (seen only by a few people); Low (barely seen by anyone).

**Height:** Height of Wind Turbines/Solar Panels. This can influence how they look in the landscape.

**Area needed to provide electricity to a house:** It is the area (in square meters) needed to install wind/solar technology to meet the electricity consumption of one Norwegian house.

**Power lines requirement:** Transmission lines are needed to connect renewable energy farms with the electricity grid. Here three categories are introduced: High (long-distance transmission lines); Medium (short-distance transmission lines); Low (near-to-zero transmission expansion needed).

# Q1: Ask ALL

| | Off-shore wind 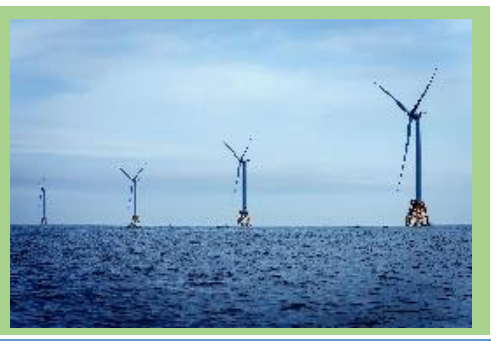 | On-shore wind 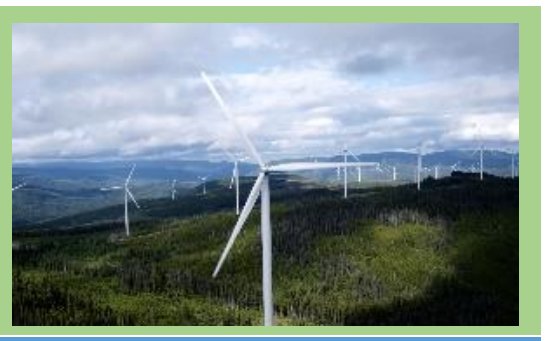 |
|---|---|---|
| Household electricity savings | 0 NOK per month | 850 NOK per month |
| Visual impact | Low | Medium to high |
| Height | Above 200 meters | Below 200 meters |
| Area needed to provide electricity to a house (m2) | 60 | 200 to 600 |
| Power lines requirement | High | Medium to high |

# Q2: If off-shore=1

| | Off-shore wind 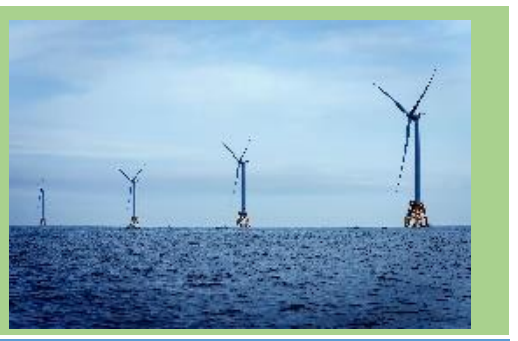 | 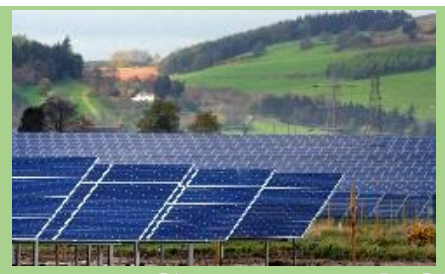 + 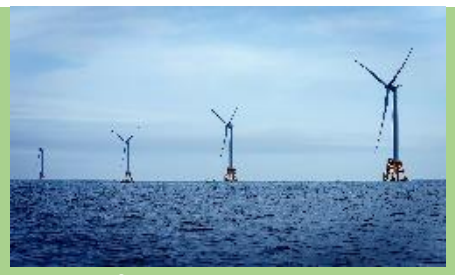 Solar + Off-shore wind |
|---|---|---|
| Household electricity savings | 0 NOK per month | 328 NOK per month |
| Visual impact | Low | Medium to Low |
| Height | Above 200 meters | Low (solar) + high (off-shore wind) |
| Area needed to provide electricity to a house ($m^2$) | 60 | 120 to150 |
| Power lines requirement | High | Medium to high |

# Q3: if on-shore=1

| | | On-shore wind large 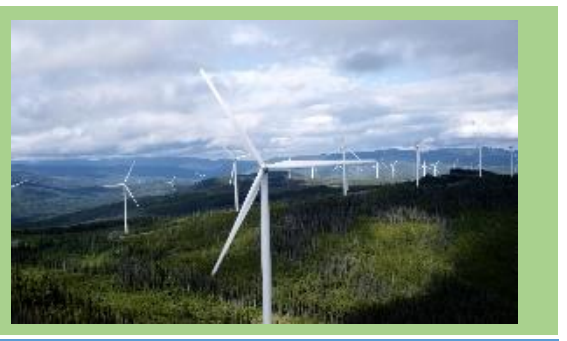 | On-shore wind small 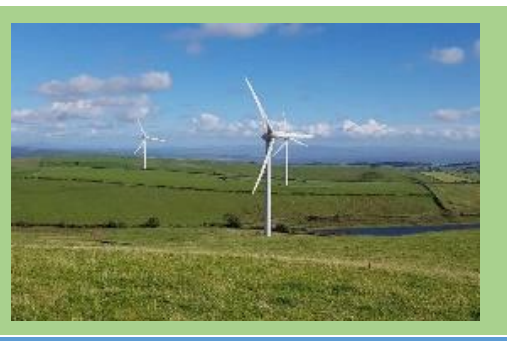 |
| --- | --- | --- | --- |
| Household electricity savings | 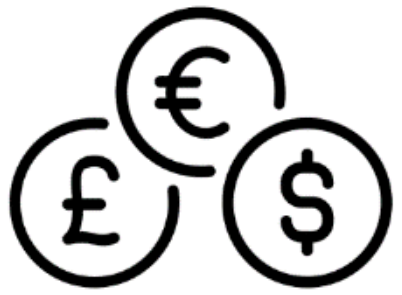 | 850 NOK per month | 1050 NOK per month |
| Visual impact | 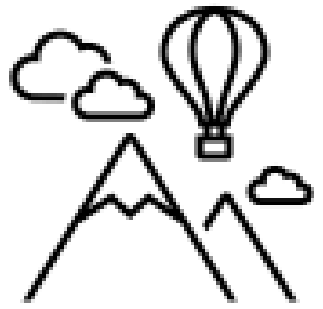 | Medium | high |
| Height | 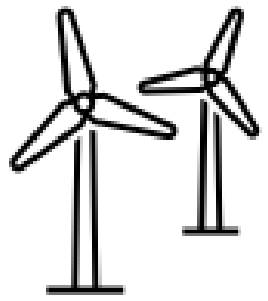 | 150-200 meters | 100-150 meters |
| Area needed to provide electricity to a house (m2) | 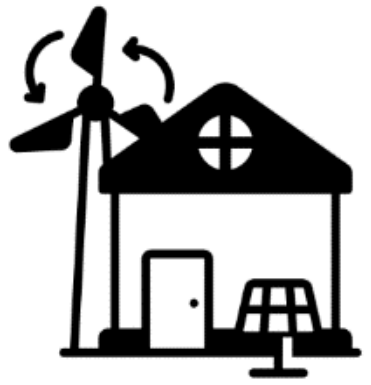 | 270 | 670 |
| Power lines requirement | 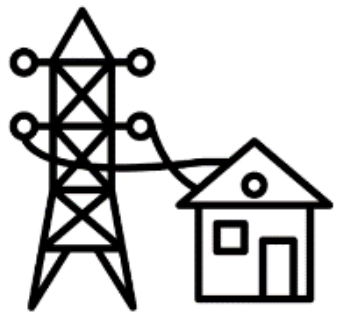 | Medium to high | Low to Medium |

# Q4: if on-shore large=1

| | On-shore wind large 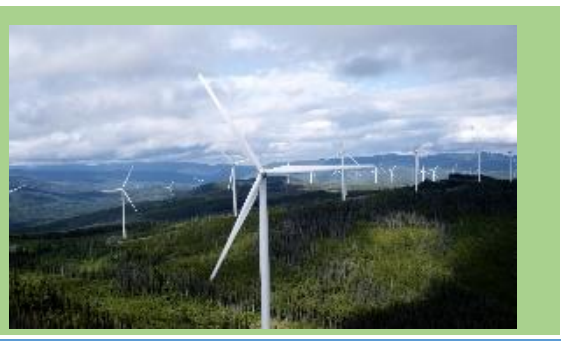 | 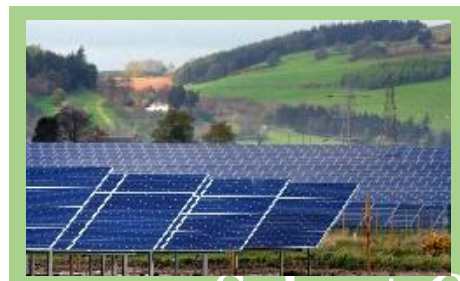 + 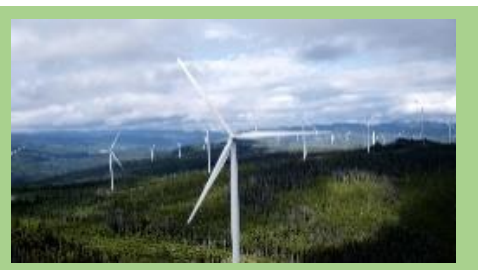 Solar + On-shore wind large |
|---|---|---|
| Household electricity savings | 847 NOK per month | 766 NOK per month |
| Visual impact | Medium | Medium |
| Height | 150-200 meters | Low (solar) + high (on-shore wind) |
| Area needed to provide electricity to a house (m2) | 270 | 310 |
| Power lines requirement | Medium to high | Medium to high |

# Q5: if on-shore small=1

| | On-shore wind small 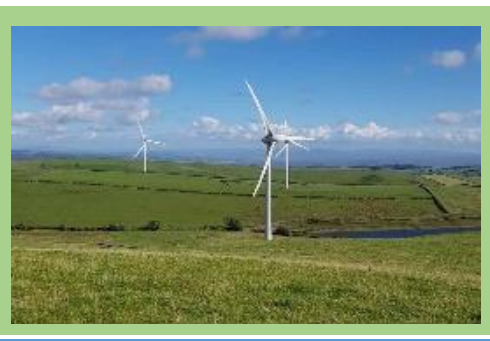 | 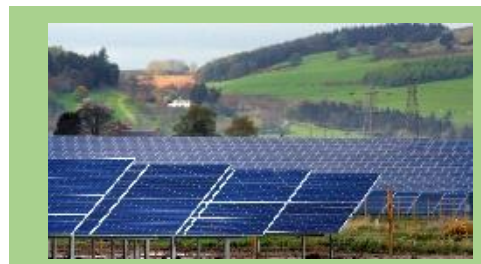 + 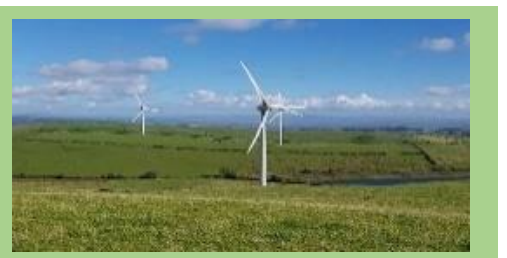 Solar + On-shore wind small |
|---|---|---|
| Household electricity savings | 847 NOK per month | 766 NOK per month |
| Visual impact | high | Medium to high |
| Height | 100-150 meters | Low (PV) + High (on-shore wind) |
| Area needed to provide electricity to a house (m2) | 670 | 410 |
| Power lines requirement | Low to Medium | Low to Medium |

**Questions for all:**

Power lines help connect new electric power stations to the local electricity grid. As Norway's electricity needs grow, we'll add more clean-energy power stations. There are two main ways to build these power lines:

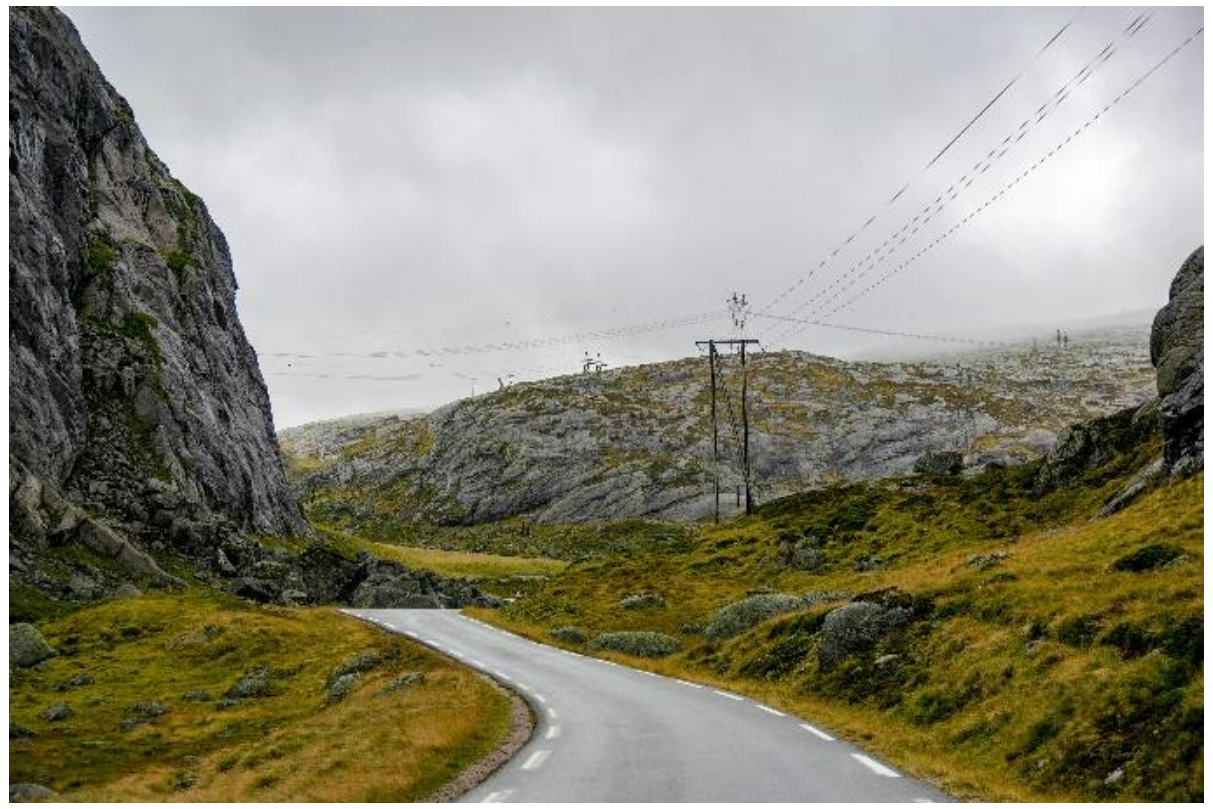

Above ground power lines

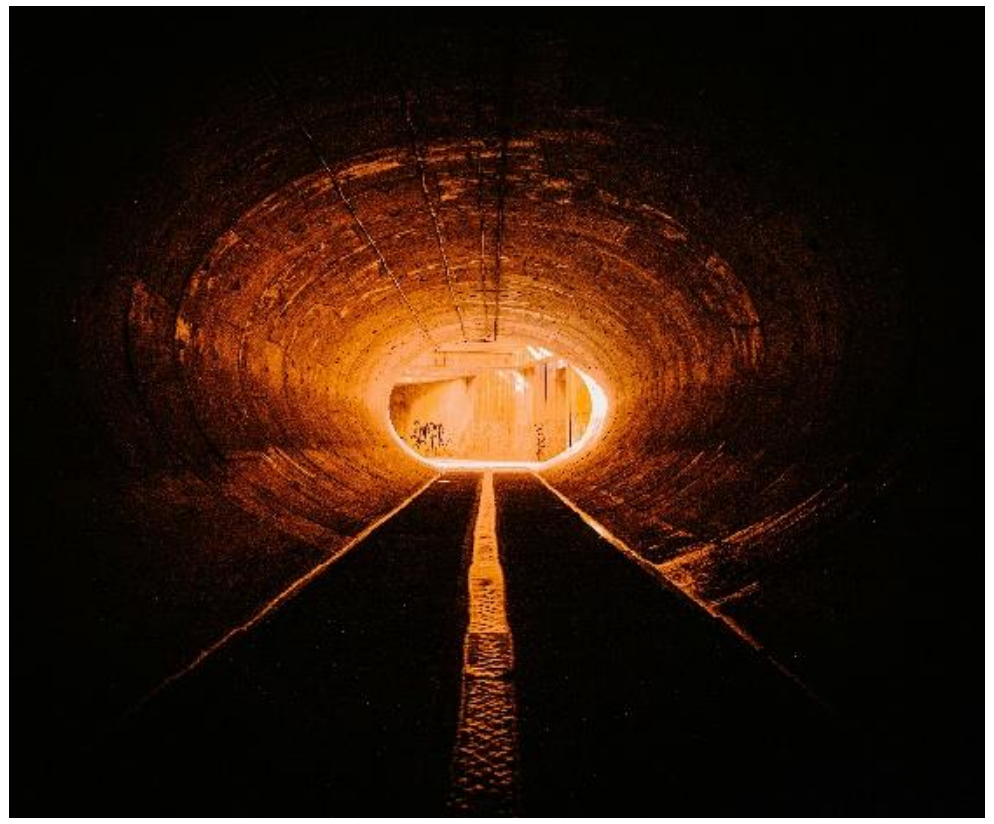

Below ground power lines
(almost 20 times more expensive)

**How would you like new power lines to be built?**

- No new power lines should be built
- Power lines above the ground
- Power lines below the ground (cost 20 times more than above-ground lines)
- Power lines below the ground in remote areas, above the ground in cities
- Power lines above the ground in remote areas, below the ground in cities

**The weather is the main driver for how much renewable energy (solar/wind) we can produce, meaning we might generate too much energy when we do not need it or not enough when we do need it. Which option do you think is best to ensure we have the right amount of energy at all times?**

- Share extra electricity with nearby countries and get extra electricity from these countries when we need it
- Keep extra electricity from renewable energy saved locally, for example, in batteries, to use later
- Use a mix approach and save of electricity locally, while also share it with nearby countries
- Try to use electricity when we make a lot of it (changing habitual patterns)

**Currently, Norway is a net electricity exporter; what do you think about the country's future energy supply?**

- Export more than it does today
- Export same as today
- Produce enough for its use (net-zero)
- Only import from other countries

**Below are some features of an energy system. Please rank them based on how important they are to you. 1 = most important**

- ○ Zero carbon emissions
- ○ Being able to produce all electricity locally from renewable energy (self-sufficiency)
- ○ Keeping the cost of electricity low
- ○ Protecting plants, animals, and birds (biodiversity)

**To what extent do you agree with the following statement?**

| | Strongly disagree | Disagree | Neither agree nor disagree | Agree | Strongly agree |
|---|---|---|---|---|---|
| Renewable energy (wind/solar) should be prioritized to meet future energy demand, even if it leads to higher electricity bills for Norwegian people. | ☐ | ☐ | ☐ | ☐ | ☐ |
| Future demand can be met by buying electricity from other countries instead of installing more power stations locally. | ☐ | ☐ | ☐ | ☐ | ☐ |
| Nature reserves (animal/bird protection) must be taken care of in deciding the wind turbine installations. | ☐ | ☐ | ☐ | ☐ | ☐ |

**Which Norwegian regions do you think are most appropriate to install wind farms (you may select more than one region):**

- ○ Agder
- ○ Innlandet
- ○ Møre og Romsdal
- ○ Nordland
- ○ Oslo
- ○ Rogaland
- ○ Vestfold og Telemark
- ○ Troms og Finmark
- ○ Trøndelag
- ○ Vestland
- ○ Viken

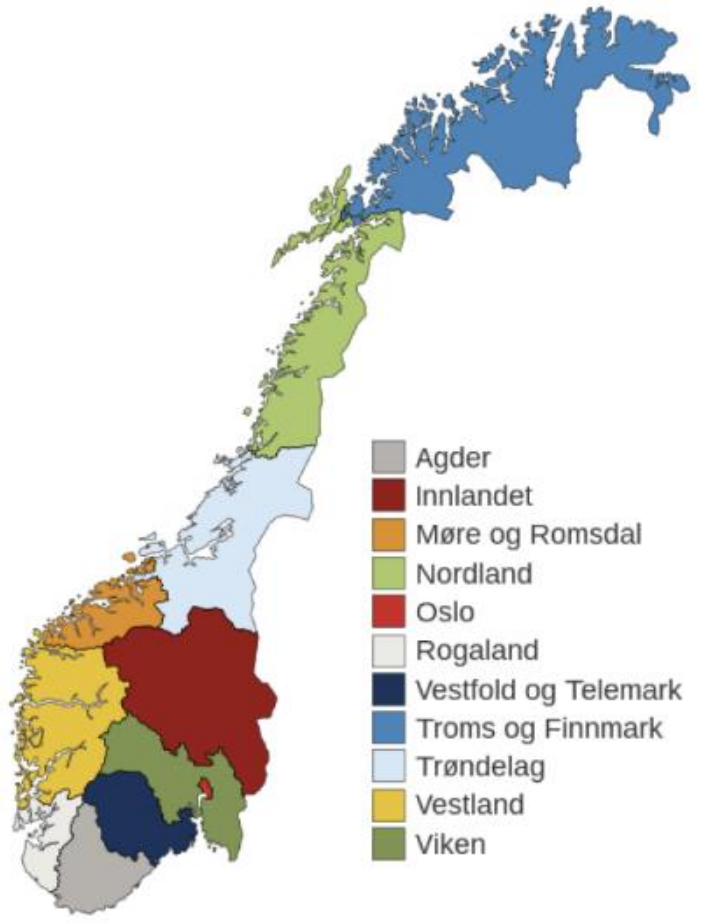

**Which Norwegian regions do you think are most appropriate to install solar farms (you may select more than one region):**

- Agder
- Innlandet
- Møre og Romsdal
- Nordland
- Oslo
- Rogaland
- Vestfold og Telemark
- Troms og Finmark
- Trøndelag
- Vestland
- Viken

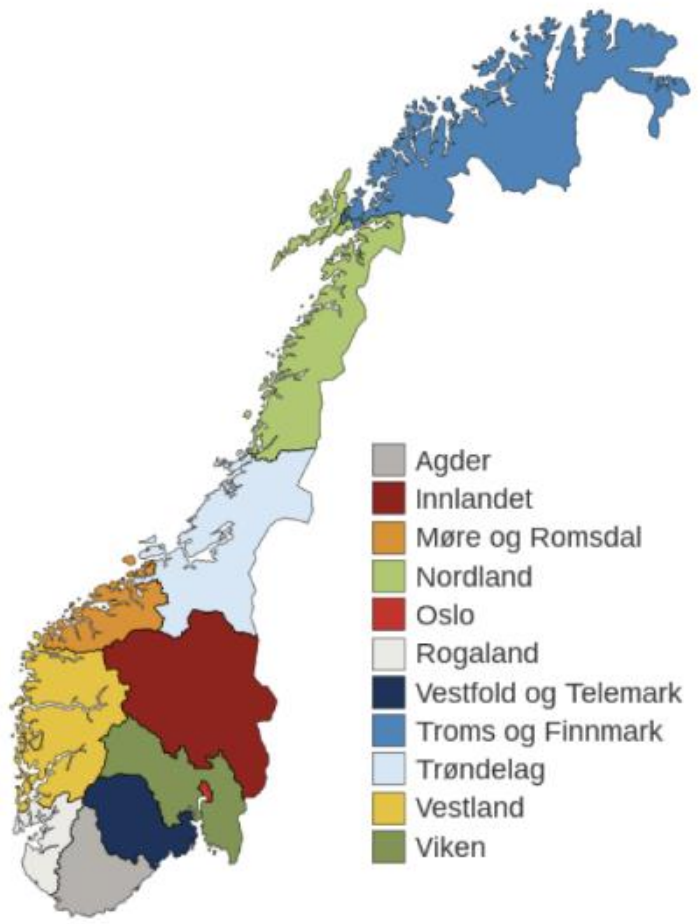

**To what extent do you agree to place wind installations in the following locations:**

| | Strongly disagree | Disagree | Neither agree nor disagree | Agree | Strongly agree |
|---|---|---|---|---|---|
| In areas where there is little economic activity to attract new industry. | ☐ | ☐ | ☐ | ☐ | ☐ |
| Near electricity consumption places where most people live, i.e., counties/municipalities? | ☐ | ☐ | ☐ | ☐ | ☐ |
| In remote areas with not much electricity consumption (high transmission cost). | ☐ | ☐ | ☐ | ☐ | ☐ |
| In areas where much of the natural landscapes are changed by human activities (cities/towns). | ☐ | ☐ | ☐ | ☐ | ☐ |
| In areas where much of the natural landscape is unaffected by human activities (forests/remote areas). | ☐ | ☐ | ☐ | ☐ | ☐ |
| Build renewable energy power stations near existing roads and power lines to minimize the need for new ones. | ☐ | ☐ | ☐ | ☐ | ☐ |

**Take a look at the pictures of different types of landscapes with and without wind turbines. To what extent do you agree or disagree to install wind farms in the following locations. Remember, there are no correct or incorrect answers. Your opinion is what counts the most.**

**Bare rock mountains:** Tall and steep landscapes with exposed rock

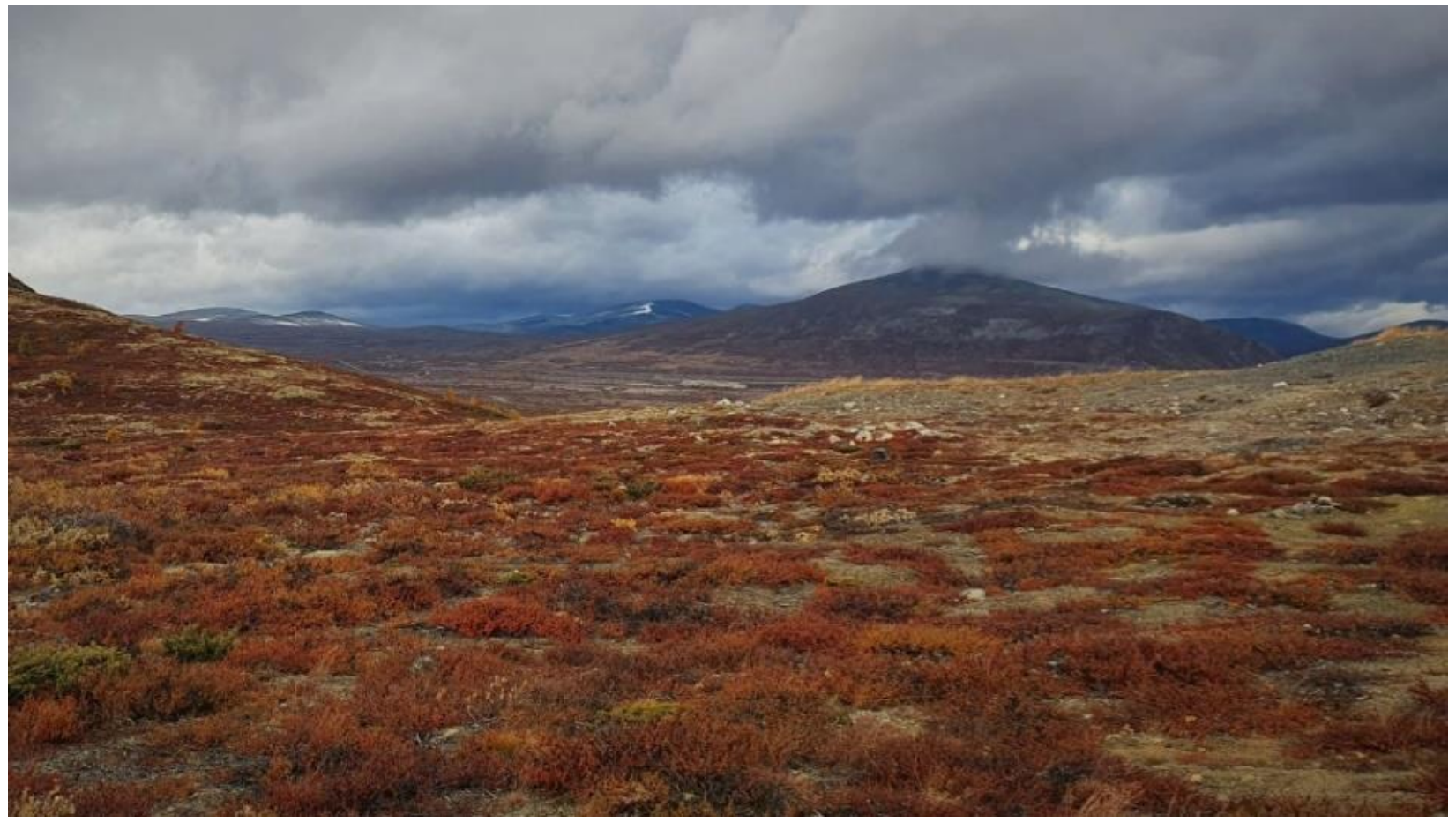

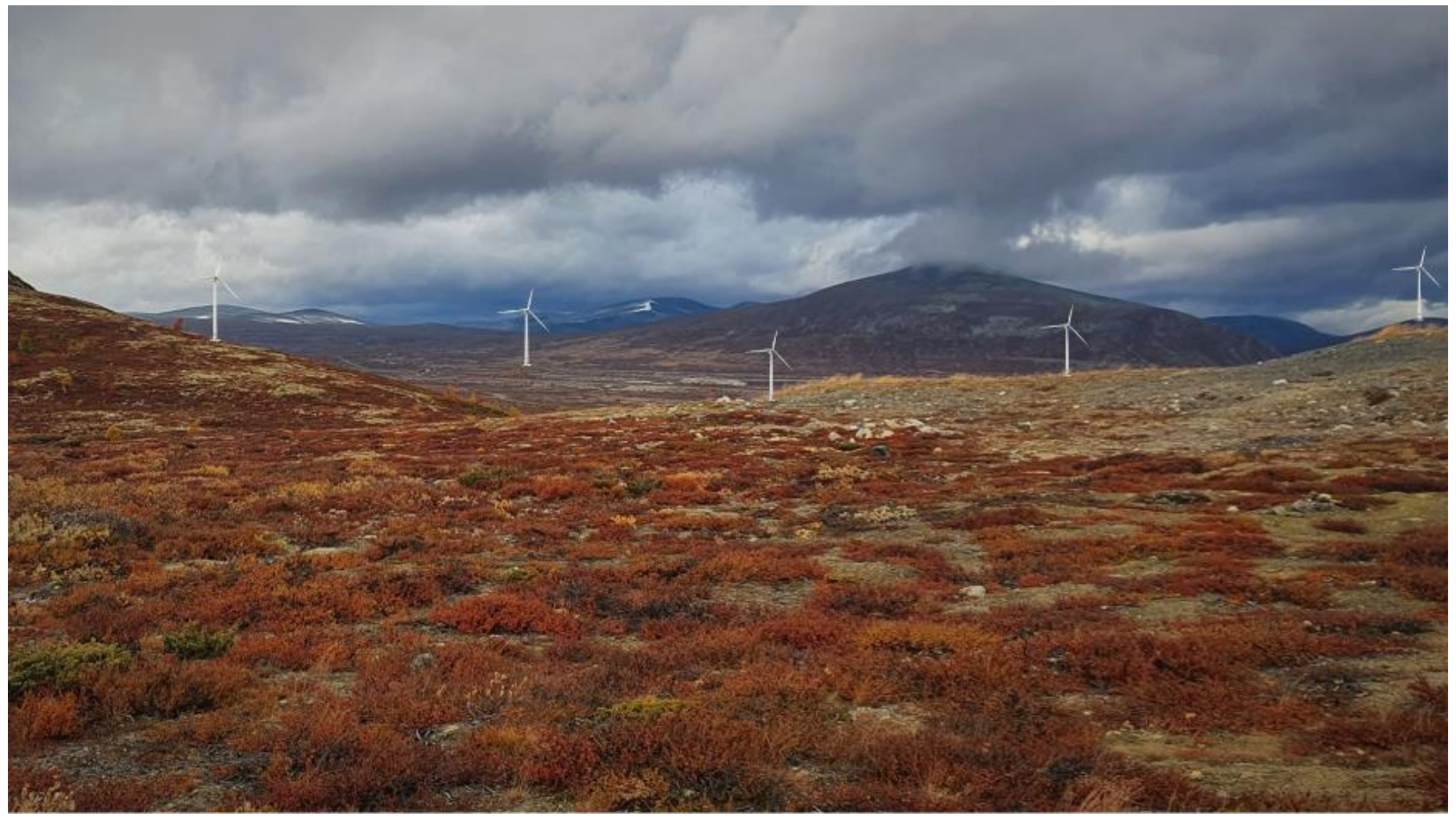

(without wind turbine installation) (With wind turbine installation)

**1: I strongly disagree, 9: strongly agree**

| Scale | **1** | **2** | **3** | **4** | **5** | **6** | **7** | **8** | **9** |
|---|---|---|---|---|---|---|---|---|---|
| Bare rock mountains | ☐ | ☐ | ☐ | ☐ | ☐ | ☐ | ☐ | ☐ | ☐ |

**Residential/Urban:** Part of a town/city where people live.

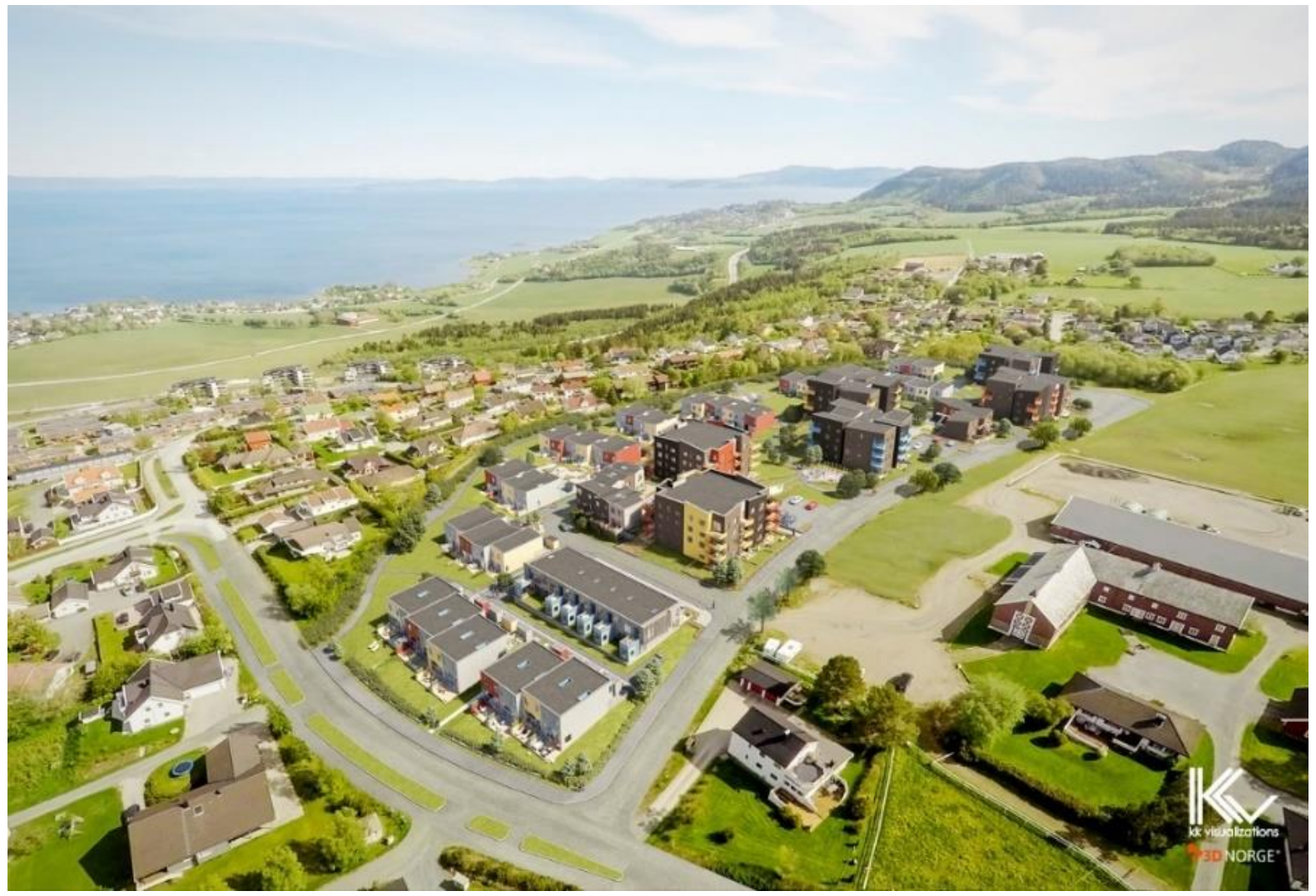

(without wind turbine installation)

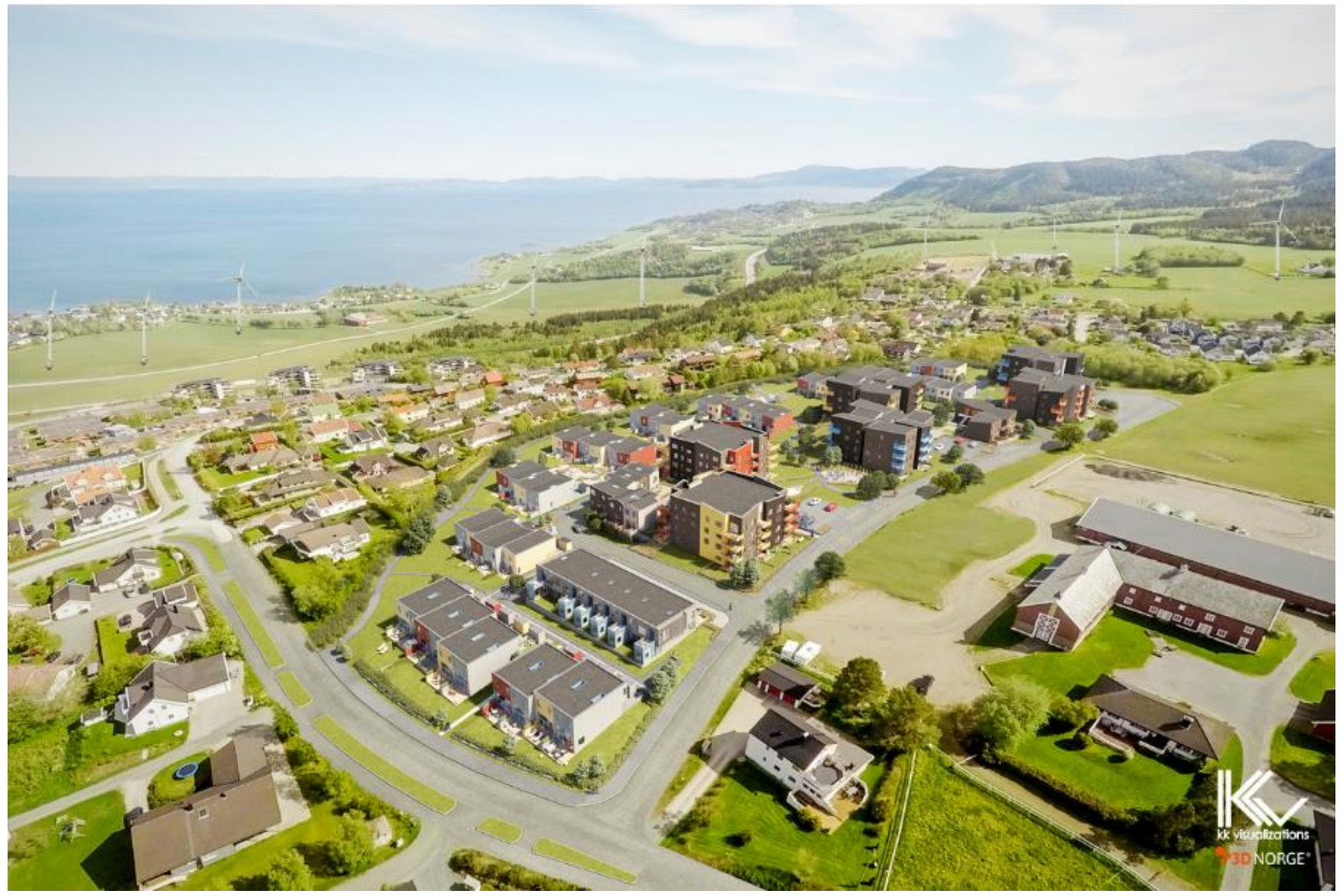

(With wind turbine installation)

**1: I strongly disagree, 9: strongly agree**

| Scale | **1** | **2** | **3** | **4** | **5** | **6** | **7** | **8** | **9** |
|---|---|---|---|---|---|---|---|---|---|
| Residential/urban | ☐ | ☐ | ☐ | ☐ | ☐ | ☐ | ☐ | ☐ | ☐ |

**Commercial/industrial zones:** area often filled with factories/warehouses where products are made/stored

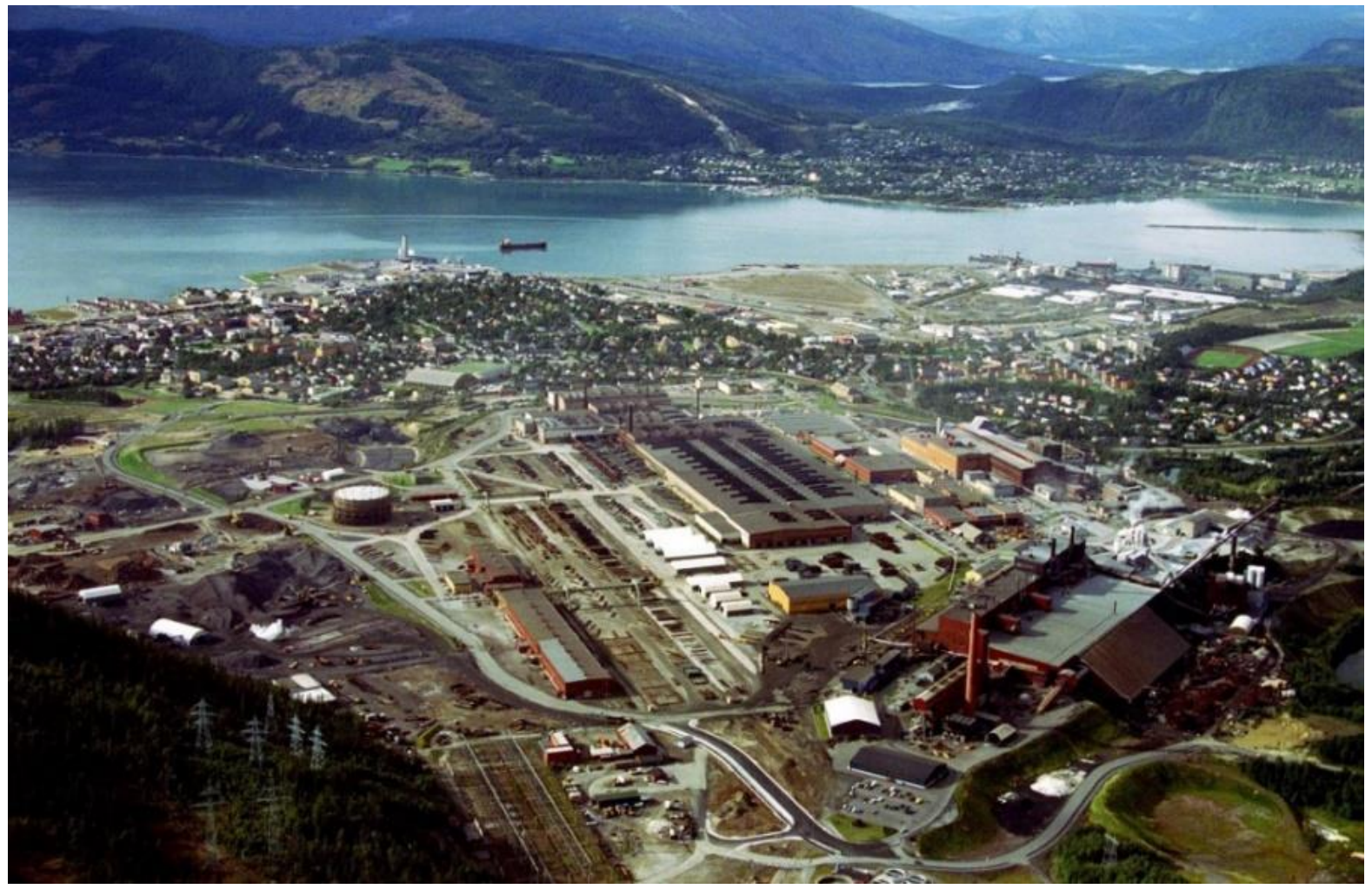

(without wind turbine installation)

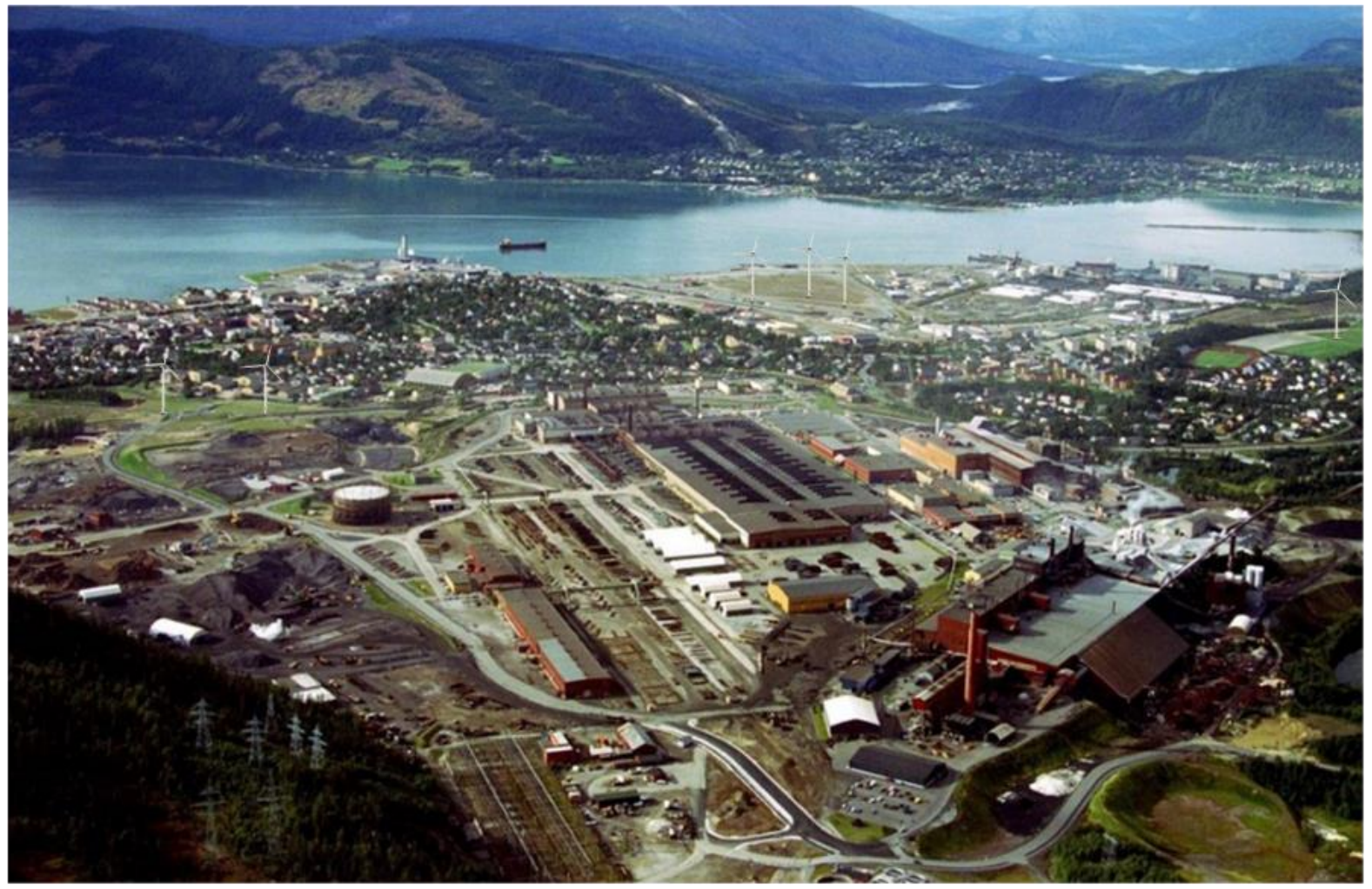

(With wind turbine installation)

**1: I strongly disagree, 9: strongly agree**

| Scale | 1 | 2 | 3 | 4 | 5 | 6 | 7 | 8 | 9 |
|---|---|---|---|---|---|---|---|---|---|
| Commercial/industrial zones | ☐ | ☐ | ☐ | ☐ | ☐ | ☐ | ☐ | ☐ | ☐ |

**Natural grassland/animal grazing land:** wide open space often covered with grasses and some trees

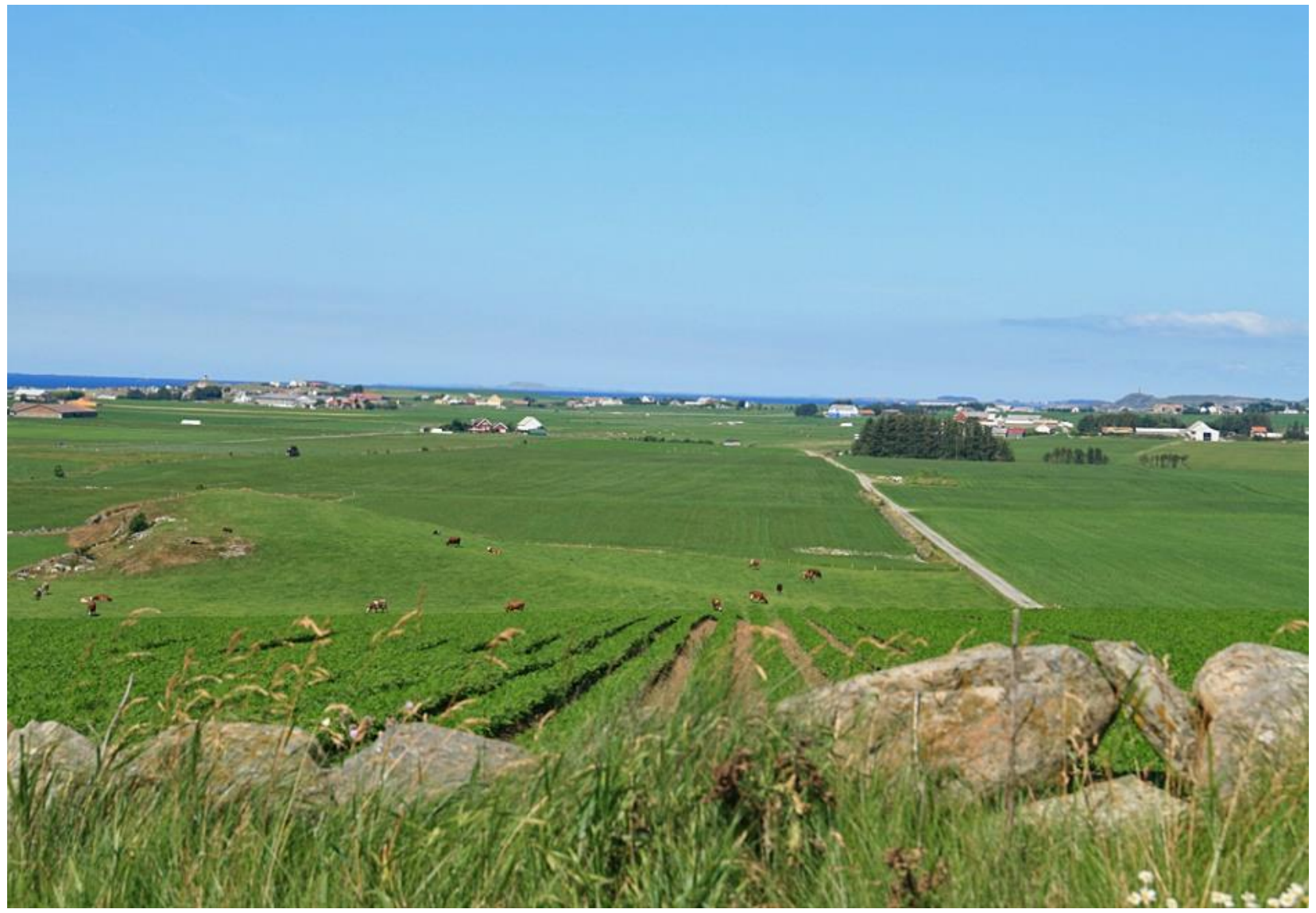

(without wind turbine installation)

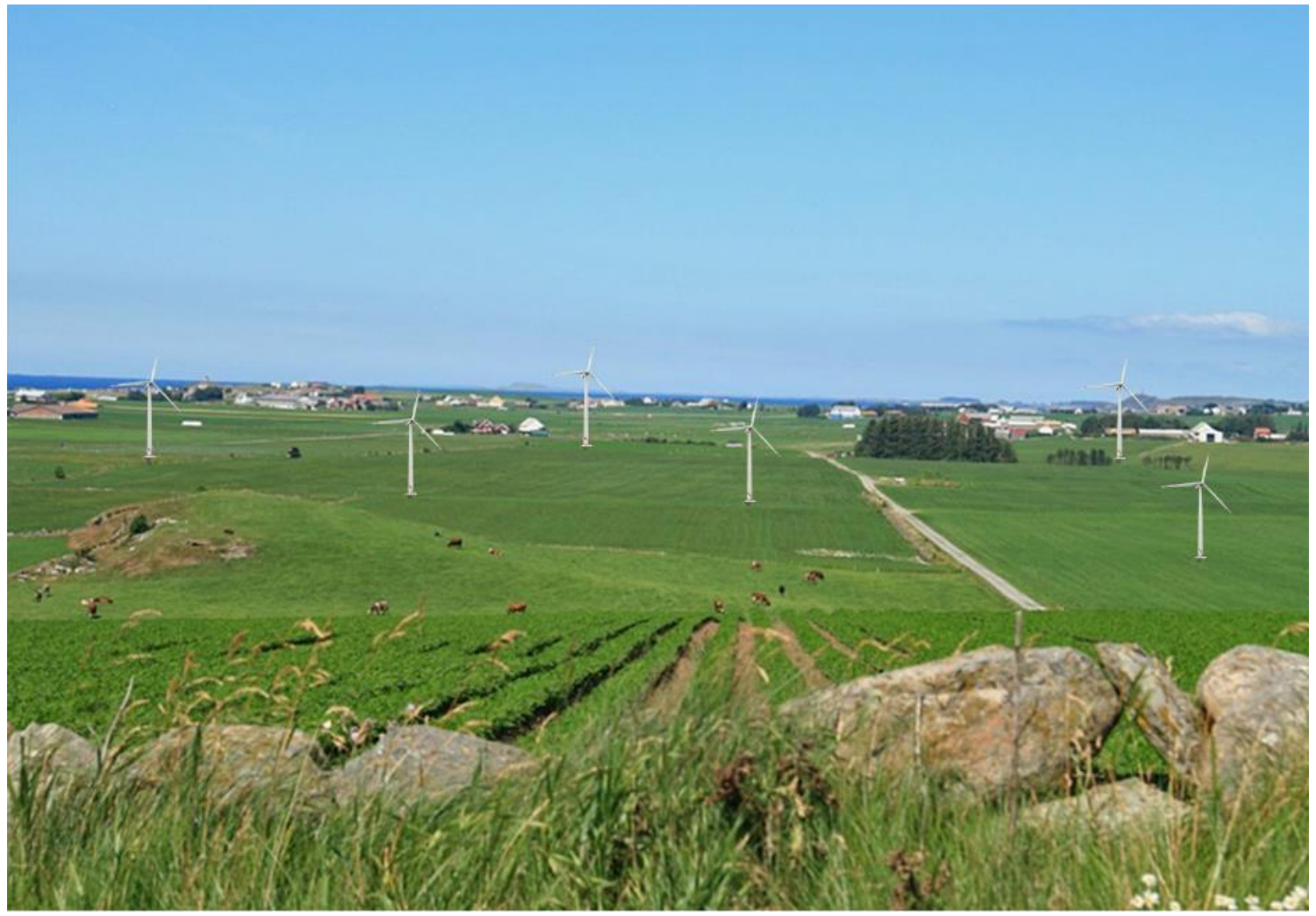

(With wind turbine installation)

**1: I strongly disagree, 9: strongly agree**

| scale | **1** | **2** | **3** | **4** | **5** | **6** | **7** | **8** | **9** |
|---|---|---|---|---|---|---|---|---|---|
| Natural grassland | ☐ | ☐ | ☐ | ☐ | ☐ | ☐ | ☐ | ☐ | ☐ |

**Nearshore areas (Beaches/dunes/coastal):** Areas near to the sea

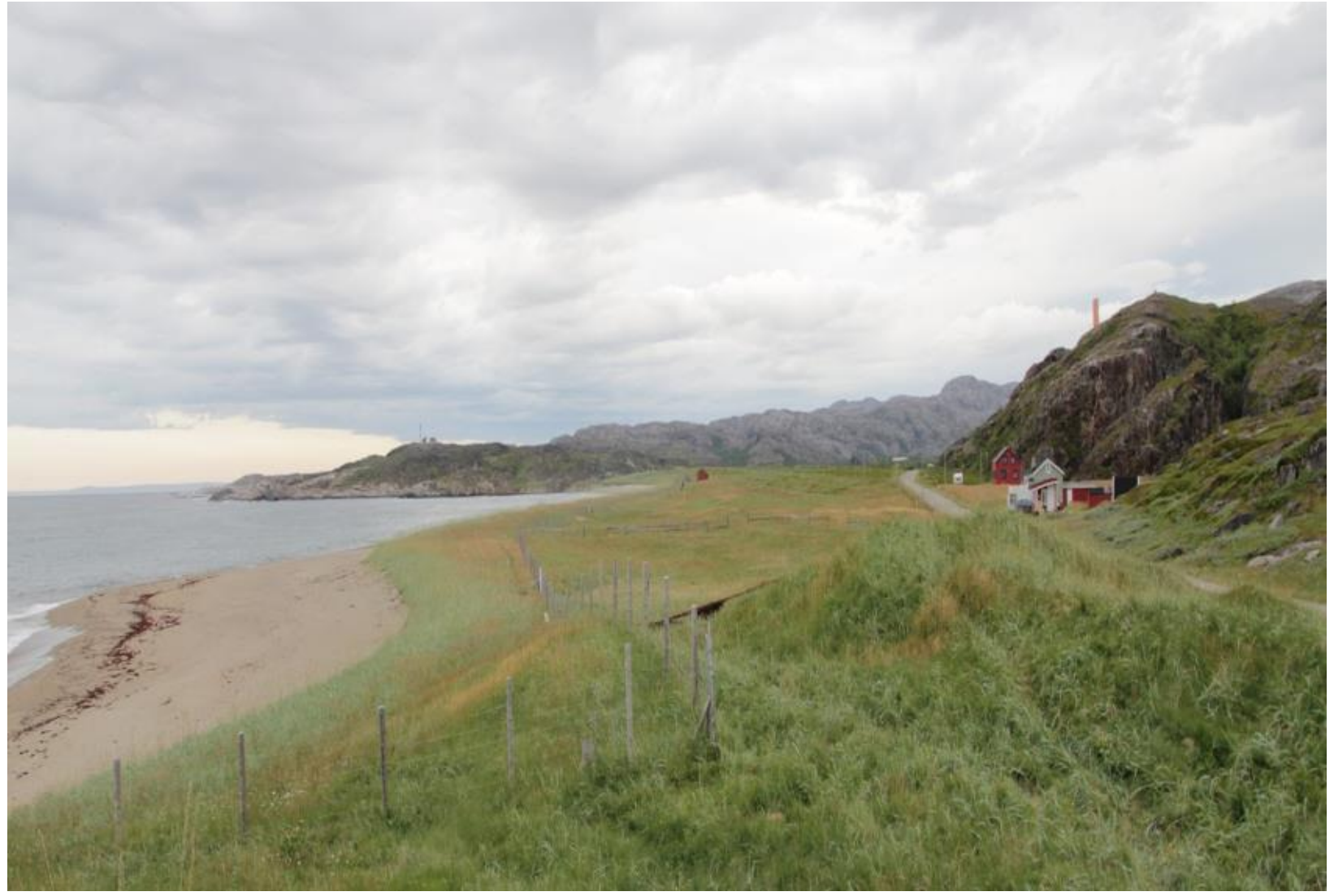

(without wind turbine installation)

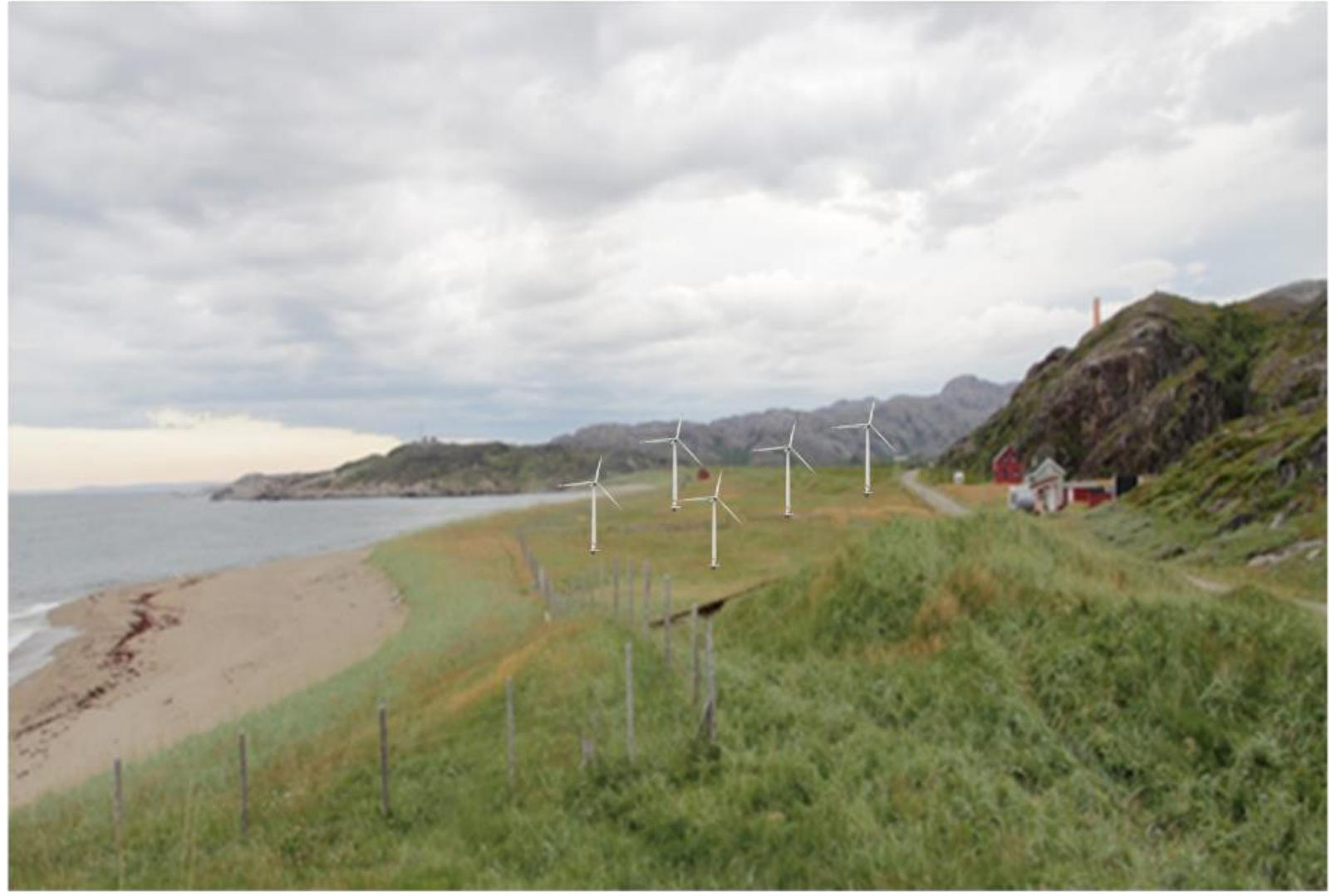

(With wind turbine installation)

**1: I strongly disagree, 9: strongly agree**

| scale | **1** | **2** | **3** | **4** | **5** | **6** | **7** | **8** | **9** |
|---|---|---|---|---|---|---|---|---|---|
| Nearshore areas | ☐ | ☐ | ☐ | ☐ | ☐ | ☐ | ☐ | ☐ | ☐ |

**Sparsely vegetated areas: Places with not many plants but patches of grass**

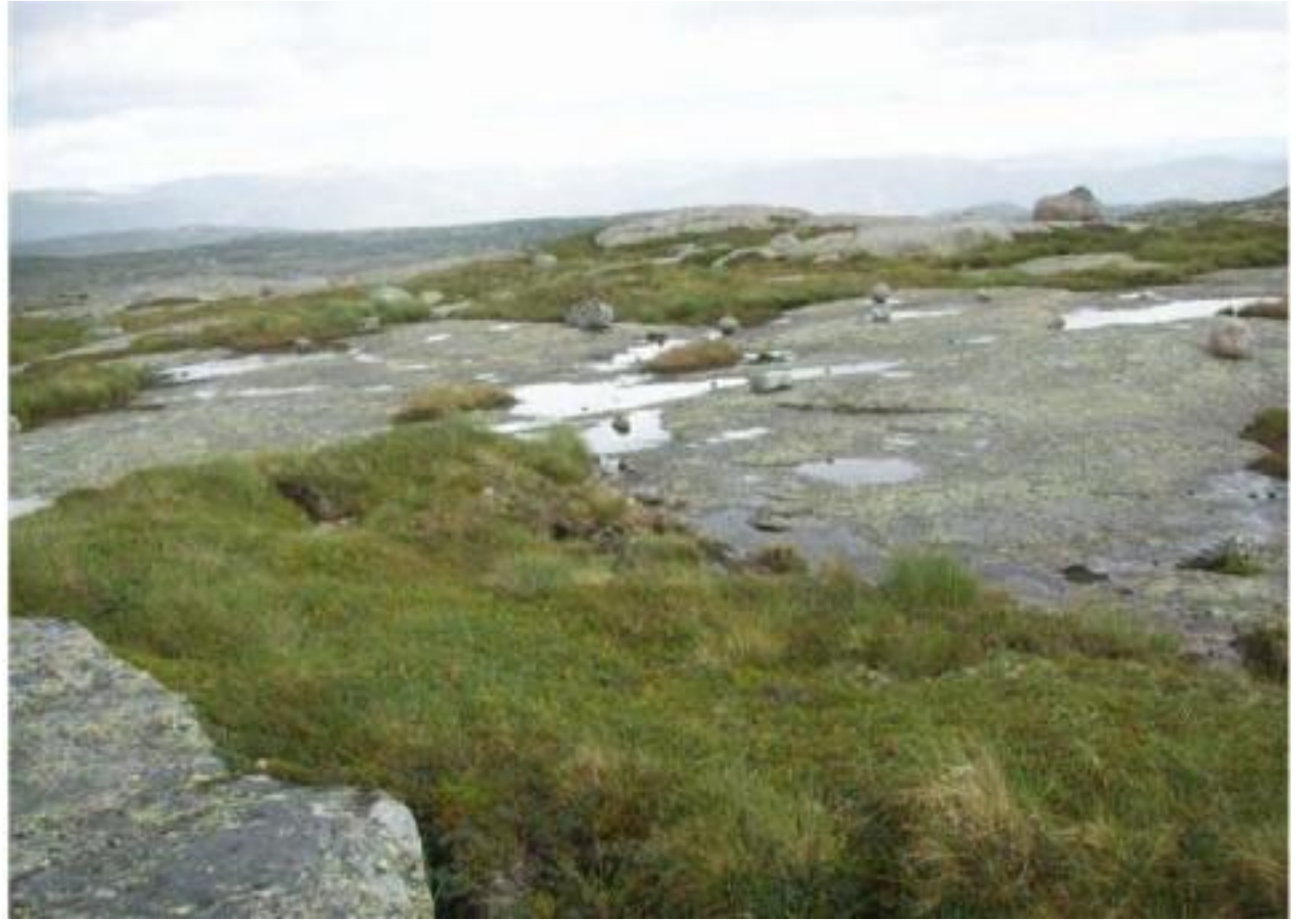

(without wind turbine installation)

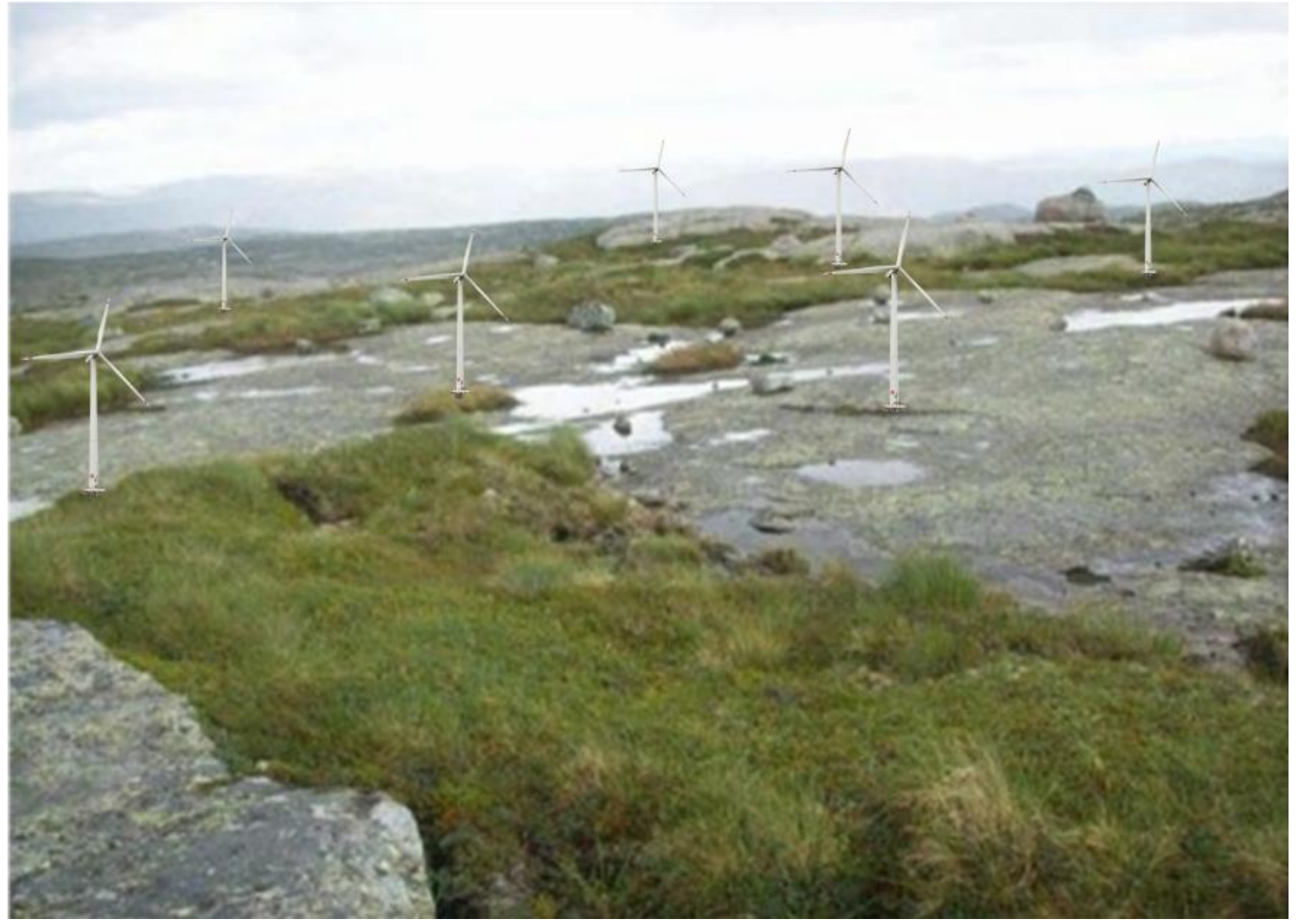

(With wind turbine installation)

**1: I strongly disagree, 9: strongly agree**

| scale | **1** | **2** | **3** | **4** | **5** | **6** | **7** | **8** | **9** |
|---|---|---|---|---|---|---|---|---|---|
| Sparsely vegetated areas | ☐ | ☐ | ☐ | ☐ | ☐ | ☐ | ☐ | ☐ | ☐ |

**Offshore areas:** a part of the ocean and far from the land (wind can be installed within a certain distance from the land, depending on how deep the water is)

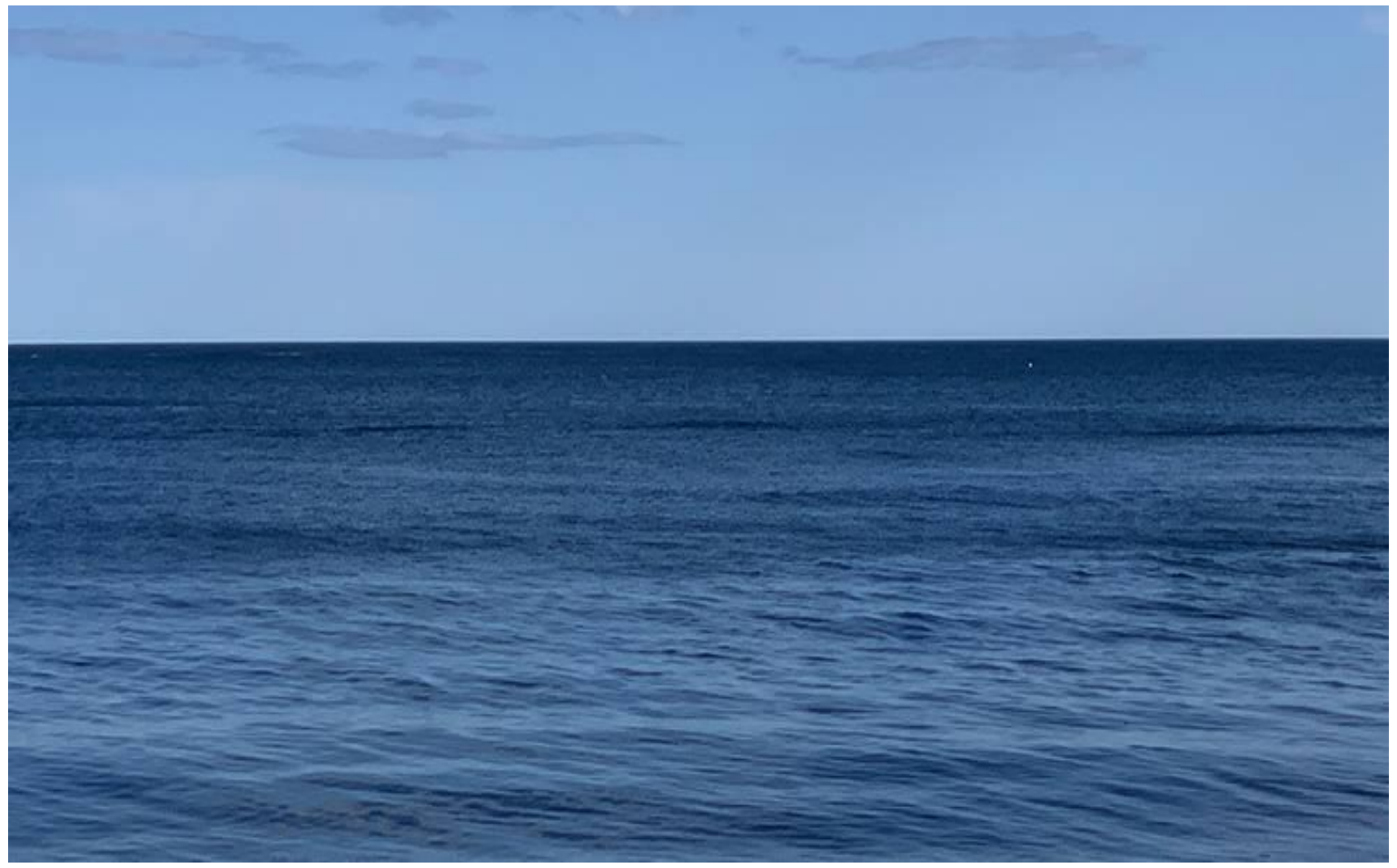

(without wind turbine installation)

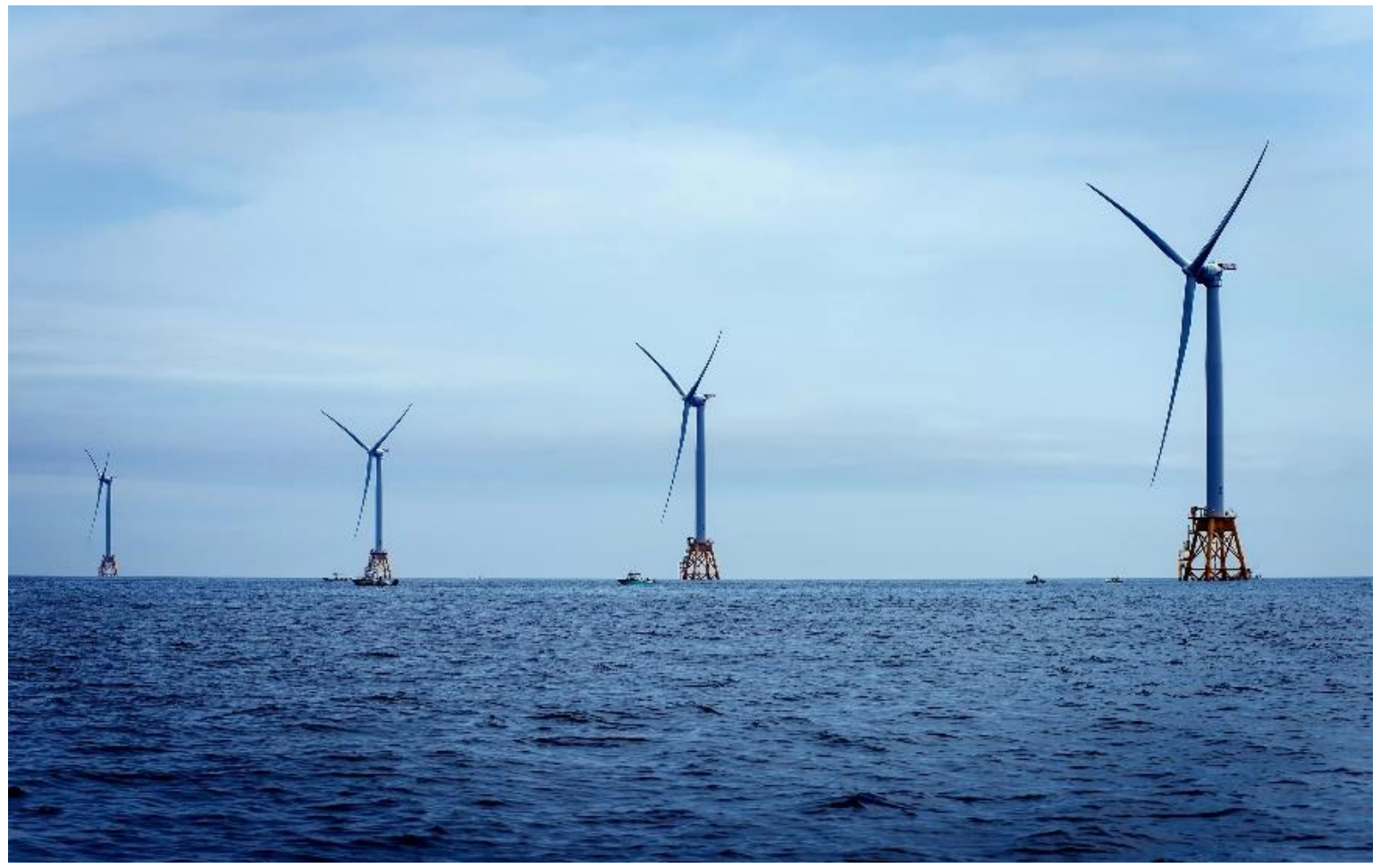

(With wind turbine installation)

**1: I strongly disagree, 9: strongly agree**

| scale | **1** | **2** | **3** | **4** | **5** | **6** | **7** | **8** | **9** |
|---|---|---|---|---|---|---|---|---|---|
| offshore areas | ☐ | ☐ | ☐ | ☐ | ☐ | ☐ | ☐ | ☐ | ☐ |

**Arable land (agriculture):** areas suitable for farming

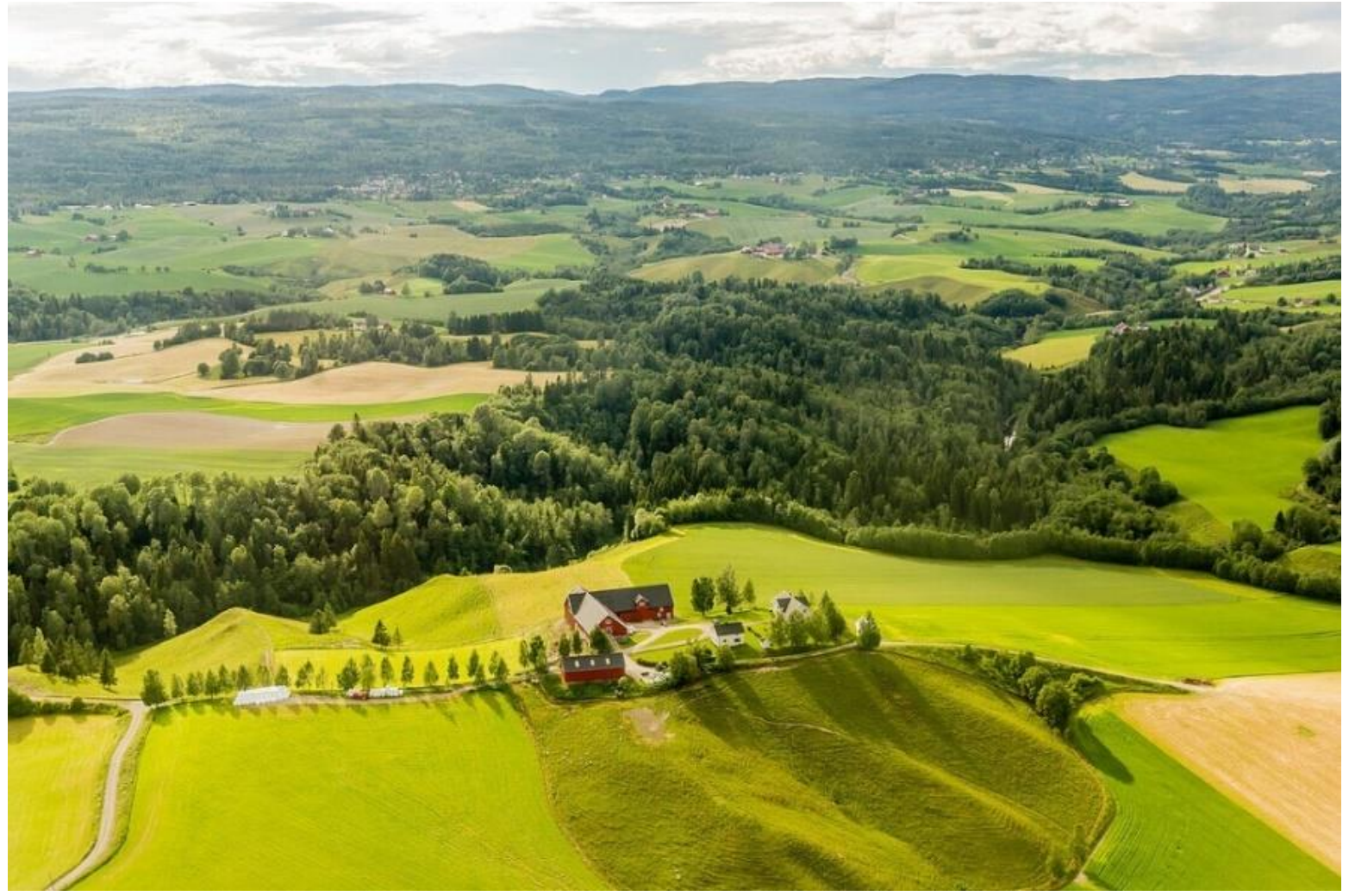

(without wind turbine installation)

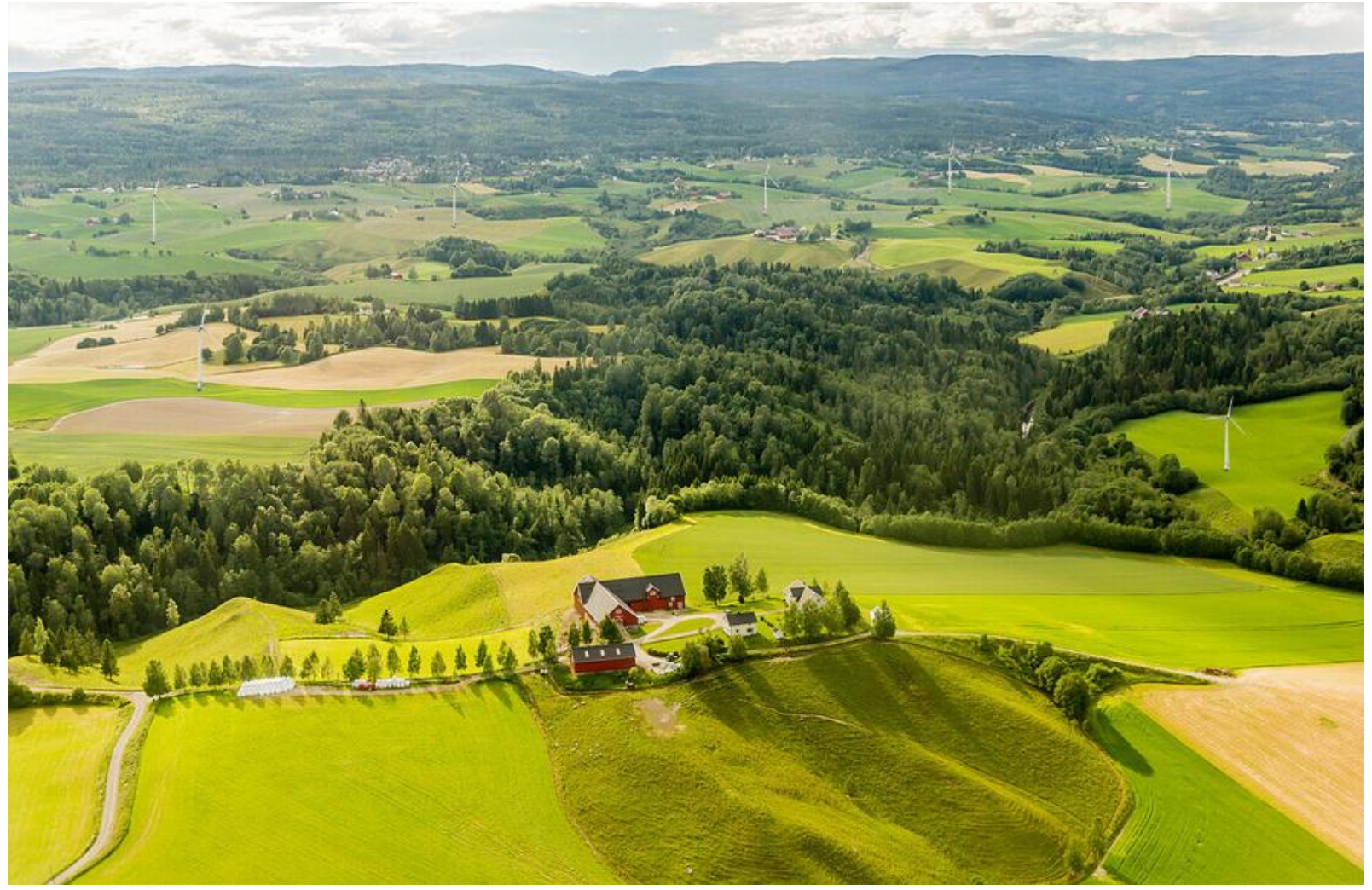

(With wind turbine installation)

**1: I strongly disagree, 9: strongly agree**

| scale | **1** | **2** | **3** | **4** | **5** | **6** | **7** | **8** | **9** |
|---|---|---|---|---|---|---|---|---|---|
| arable land | ☐ | ☐ | ☐ | ☐ | ☐ | ☐ | ☐ | ☐ | ☐ |

**Forests (Mixed type):** large areas covered with different types of trees but also have patches of land

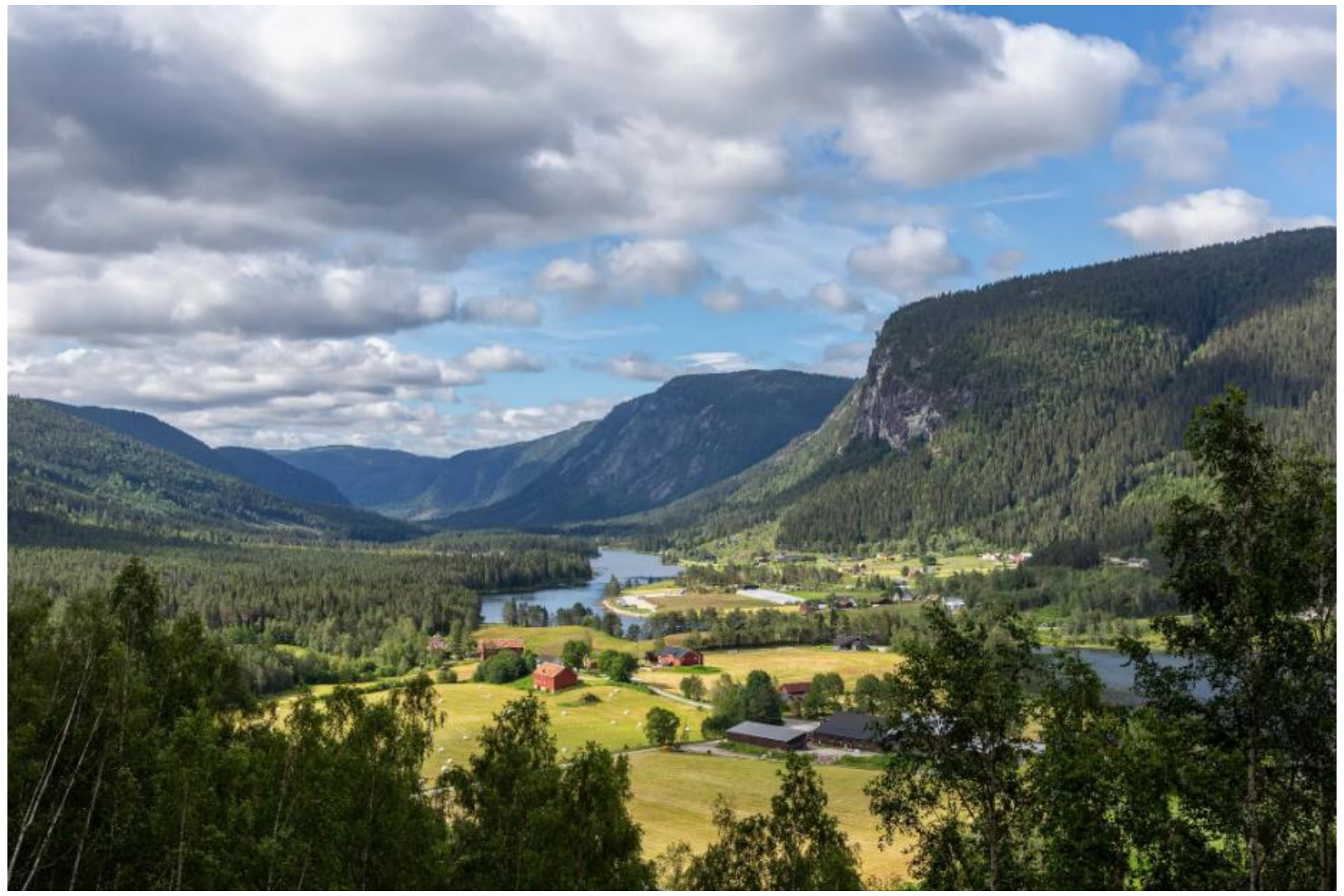

(without wind turbine installation)

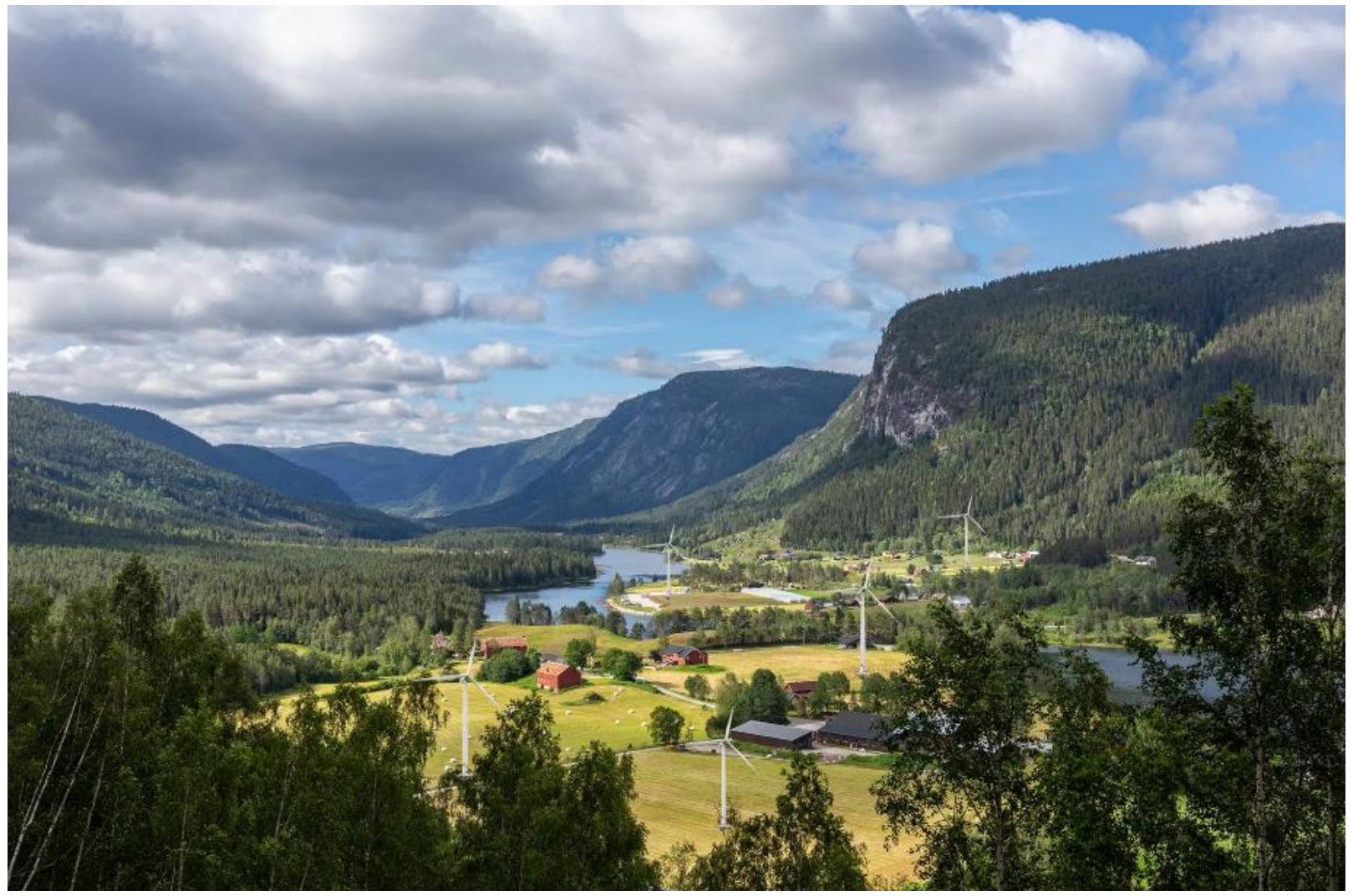

(With wind turbine installation)

**1: I strongly disagree, 9: strongly agree**

| scale | **1** | **2** | **3** | **4** | **5** | **6** | **7** | **8** | **9** |
|---|---|---|---|---|---|---|---|---|---|
| Forests (Mixed type) | ☐ | ☐ | ☐ | ☐ | ☐ | ☐ | ☐ | ☐ | ☐ |